\documentclass[aps,pra,10pt,a4paper,superscriptaddress,onecolumn,preprintnumbers,floatfix,nofootinbib, showkeys]{revtex4-2}

\usepackage{graphicx}
\usepackage[utf8]{inputenc}
\usepackage{tocloft}
\usepackage{epsf}
\usepackage{bm}
\usepackage{amsmath}
\usepackage{amsfonts}
\usepackage{amssymb}
\usepackage{color}
\usepackage{caption}
\usepackage{subcaption}
\usepackage{booktabs} 
\usepackage{float}
\usepackage{empheq}

\usepackage[citecolor=blue,urlcolor=blue,unicode=true,pdfusetitle, bookmarks=true,bookmarksnumbered=true,bookmarksopen=true,bookmarksopenlevel=3, breaklinks=false,pdfborder={0 0 1},backref=false,colorlinks=true, linkcolor=blue]{hyperref}

\numberwithin{equation}{section}

\makeatletter

\renewcommand*{\p@subsection}{}

\renewcommand*{\p@subsubsection}{}

\renewcommand*{\p@paragraph}{}
\makeatother

\providecommand{\U}[1]{\protect\rule{.1in}{.1in}}

\newcommand{\be}{\begin{equation}}
\newcommand{\ee}{\end{equation}}

\newcommand{\mincir}{\raise
-3.truept\hbox{\rlap{\hbox{$\sim$}}\raise4.truept\hbox{$<$}\ }}
\newcommand{\magcir}{\raise
-3.truept\hbox{\rlap{\hbox{$\sim$}}\raise4.truept\hbox{$>$}\ }}

\begin{document}
\title{Crunching, Bouncing, and Cyclical Cosmologies from Dark Sector Interactions}

\author{Marcel van der Westhuizen}
\email{marcelvdw007@gmail.com}
\affiliation{Centre for Space Research, North-West University, Potchefstroom 2520, South
Africa}
\author{Amare Abebe}
\email{amare.abebe@nithecs.ac.za}
\affiliation{Centre for Space Research, North-West University, Potchefstroom 2520, South
Africa}
\affiliation{National Institute for Theoretical and Computational Sciences (NITheCS),
South Africa}

\begin{abstract}

We present new mechanisms that produce either a future Big Crunch turnaround or a past non-singular bounce in flat FLRW cosmologies within general relativity at the background level, driven solely by non-gravitational interactions between dark matter (DM) and dark energy (DE). We study phenomenological interacting dark energy (IDE) models based on linear kernels of the form $Q = 3H(\delta_{\rm dm}\rho_{\rm dm} + \delta_{\rm de}\rho_{\rm de})$, focusing on parameter regimes with strong energy transfer from dark energy to dark matter. In this strong interacting regime, the interaction does not vanish when one component crosses zero density, allowing one of the dark-sector densities to become negative. The resulting sign changes can violate the energy conditions required for cosmological turnarounds in a flat universe, thereby enabling either (i) a maximum scale factor followed by recollapse into a big crunch, or (ii) a minimum non-zero scale factor corresponding to a bounce. We derive analytic conditions for these turnarounds and obtain closed-form expressions for the associated maximum or minimum scale factor. We also show that, in a closed universe, a special case of the same IDE framework can be tuned to yield a cyclic scenario. Although these strong interaction scenarios are unlikely to describe the observed Universe, they provide a concrete demonstration that exotic cosmological behaviour can arise naturally in underexplored regions of the parameter space of familiar IDE models.
\end{abstract}
\keywords{Interacting Dark Energy; Big Bounce; Big Crunch; Cyclic Cosmology;  Negative Energy}\date{\today}
\maketitle
\date{\today }

\section{Introduction}

The standard $\Lambda$CDM model describes a universe governed by general relativity in a spatially flat Friedmann--Lemaître--Robertson--Walker (FLRW) spacetime. Its energy content consists of radiation, baryons, cold dark matter (DM), and dark energy (DE), the latter commonly modelled as a cosmological constant $\Lambda$ responsible for the observed late-time accelerated expansion of the universe, which is expected to persist indefinitely into the future. In the early universe, $\Lambda$CDM further assumes a period of rapid accelerated expansion known as inflation, possibly preceded by a Big Bang singularity. The $\Lambda$CDM model is well supported by both early-time CMB data \cite{Planck:2018vyg} and late-time supernovae observations \cite{Brout:2022vxf,DES:2025sig}, but theoretical problems remain, and improved measurements from other cosmological probes have questioned the validity of the standard model. 

Theoretical concerns related to early-time cosmology include the problem of geodesic incompleteness at the Big Bang singularity and the fine tuning required to make models of inflation work \cite{Borde:2001nh, Ijjas:2018qbo}. 
Cosmological models that avoid either a past singularity or an inflationary phase typically do so by introducing a bounce preceded by a period of contraction. Well-studied examples include ekpyrotic scenarios with colliding branes \cite{Khoury:2001wf,Khoury:2001bz,Khoury:2003rt,Cai:2012va}, string-inspired constructions \cite{Gasperini:1992em,Biswas:2005qr,Biswas:2006bs}, matter-bounce models \cite{Finelli:2001sr,Cai:2007zv,Brandenberger:2010dk,Cai:2011zx}, quantum-inspired bounces \cite{Peter:2008qz, Gingrich:2026ghf, Lustosa:2025hgl}, and modified gravity frameworks such as Loop Quantum Gravity \cite{Bojowald:2001xe, Cai:2014zga, Odintsov:2015uca,  Schander:2015eja, Amoros:2013nxa} and higher-order curvature theories \cite{Brandenberger:1993ef, Carloni:2005ii, Cai:2011tc, Nojiri:2014zqa, Pavlovic:2017umo,  Sahoo:2019qbu, Ilyas:2020qja, Ageeva:2021yik}. Reviews on bouncing cosmology are given in \cite{Novello:2008ra, Battefeld:2014uga, Brandenberger:2016vhg}. This is still an active area of discussion, as seen by recent examples of models featuring non-singular bounces or even cyclical behaviour \cite{Chokyi:2026xwi,Rani:2026uxv,  Benetti_2026, Mukherjee:2026pym, Zhang:2026cux, Sherpa:2026bzw, Vitenti:2026hgs, Stingley:2025uks, Ahmed:2025mkn, Benetti:2025mmn, Limongi:2025jsh,  Ling:2025ncw, Itzhaki:2025gdv, Singh:2023gxd, Bojowald:2025ocr, Asenjo:2025xtj,Lymperis:2025fbs}.

Beyond these early-time concerns, a number of challenges are associated with late-time expansion and the dark sector. These include the coincidence problem \cite{Amendola:1999er,Zimdahl:2001ar,Chimento:2003iea,Farrar:2003uw,Wang:2004cp,Olivares:2006jr,Sadjadi:2006qp,Quartin:2008px,delCampo:2008jx,Caldera-Cabral:2008yyo,He:2010im,delCampo:2015vha}, the cosmological constant problem \cite{Weinberg:1988cp}, the $H_0$ and $S_8$ tensions \cite{Verde:2019ivm,DiValentino:2020zio,DiValentino:2021izs,Perivolaropoulos:2021jda,Schoneberg:2021qvd,Shah:2021onj,Abdalla:2022yfr,DiValentino:2022fjm,Kamionkowski:2022pkx,Giare:2023xoc,Hu:2023jqc,Verde:2023lmm,DiValentino:2024yew,CosmoVerseNetwork:2025alb,Ong:2025cwv}, and the reported preference for dynamical dark energy in recent baryon acoustic oscillation measurements \cite{DESI:2025zgx,DESI:2025fii,DESI:2025qqy,DESI:2025wyn}. These issues have motivated a broad range of alternatives, including modified gravity \cite{DiValentino:2015bja,Zumalacarregui:2020cjh,Odintsov:2020qzd,DeFelice:2020cpt,Pogosian:2021mcs,CANTATA:2021asi,Schiavone:2022wvq,Ishak:2024jhs,Specogna:2023nkq,AtacamaCosmologyTelescope:2025nti} and dynamical dark energy \cite{DESI:2024mwx,Cortes:2024lgw,Luongo:2024fww,Gialamas:2024lyw,Dinda:2024kjf,Bhattacharya:2024hep,Giare:2024gpk,Li:2024qus,Paliathanasis:2025cuc, Vasquez:2025enh, Li:2025vuh, Xu:2026sbw}. In this paper, we restrict attention to non-gravitational interactions within the dark sector \cite{Kumar:2016zpg,Murgia:2016ccp,Kumar:2017dnp,DiValentino:2017iww,Kumar:2021eev,Gao:2021xnk,Pan:2023mie,Benisty:2024lmj,Yang:2020uga,Forconi:2023hsj,Pourtsidou:2016ico,DiValentino:2020vnx,DiValentino:2020leo,Nunes:2021zzi,Yang:2018uae,vonMarttens:2019ixw,Lucca:2020zjb,Gao:2022ahg,Zhai:2023yny,Bernui:2023byc,Becker:2020hzj,Hoerning:2023hks,Giare:2024ytc,Escamilla:2023shf,vanderWesthuizen:2023hcl,Silva:2024ift,DiValentino:2019ffd,Li:2024qso,Pooya:2024wsq,Halder:2024uao,Castello:2023zjr,Yao:2023jau,Mishra:2023ueo,Nunes:2016dlj,Yang:2025uyv,Paliathanasis:2026ymi,Millano:2026hjk, Dai:2026pvx, vanderWesthuizen:2025I,vanderWesthuizen:2025II,vanderWesthuizen:2025III}. For review papers on dark sector interactions, see~\cite{vanderWesthuizen:2025III, Bolotin:2013jpa, Wang:2016lxa, Wang:2024vmw}.

Interacting dark energy (IDE) models were originally proposed as phenomenological extensions of $\Lambda$CDM that could alleviate one or more of the preceding late-time issues, depending on the interaction kernel and the datasets considered. Of special recent interest is that dark sector interactions may provide a mechanism for dynamical dark energy preferred by DESI and could allow for a phantom crossing of the effective dark energy equation of state \cite{Guedezounme:2025wav, Silva:2025hxw,Pan:2025qwy,Shah:2025ayl,Lee:2025pzo,vanderWesthuizen:2025iam, Giare:2024smz,Zhu:2025lrk,Gonzalez-Espinoza:2025vrc,MV:2025yjt,Lyu:2025nsd,Wu:2025vrl,Li:2025ula, Li:2025muv, Li:2026xaz, Figueruelo2026IDEconstraints, Turyshev:2026ewm}. At the same time, IDE models may introduce their own difficulties, including the absence of a unique fundamental field-theory description (though attempts at this include \cite{He:2011qn, Lucca_2020, Johnson:2020gzn, Pan:2020zza, Aboubrahim:2024cyk}), and the appearance of both potentially unstable regions \cite{M.B.Gavela_2009,Honorez_2010} and negative energy densities in parts of parameter space \cite{Caldera_Cabral_2009_DSA,Caldera_Cabral_2009_structure,V_liviita_2010,von_Marttens_2020,vanderWesthuizen:2025I,vanderWesthuizen:2025II,vanderWesthuizen:2025III}.  Nevertheless, IDE remains a well-studied framework whose less explored regimes can exhibit qualitatively new cosmological behaviour. 

In this work, we focus on a class of interacting dark energy models that are usually regarded as pathological and therefore largely ignored, namely those regions of parameter space in which one of the dark-sector energy densities becomes negative, identified in \cite{vanderWesthuizen:2025I}. Rather than discarding these regimes \emph{a priori}, we study their background dynamics at the homogeneous level and show that they can naturally give rise to cosmological turnarounds. Specifically, we demonstrate that strong non-gravitational energy transfer between dark energy and dark matter can generate either a future big-crunch singularity, a past non-singular bounce, or, in the presence of spatial curvature, a cyclic cosmological scenario, without modifying general relativity.

Much of the IDE literature focuses on future big-rip singularities \cite{Nojiri_2005_IDE,Curbelo_2006,Timoshkin:2010zz,Pan_2020,deHaro:2023lbq,vanderWesthuizen:2023hcl,vanderWesthuizen:2025I,vanderWesthuizen:2025II,vanderWesthuizen:2025III}, while comparatively little attention has been given to future big-crunch singularities or to avoiding an initial big-bang singularity through a non-singular bounce. For broader discussions of cosmological singularities, see \cite{Hawking:1970zqf, Trivedi:2023zlf}, while a review with emphasis on big rip and bouncing scenarios, including brief sections in IDE contexts, is given in \cite{deHaro:2023lbq} and \cite{Cai:2016hea}, respectively.\cite{Cai:2016hea}. 

A common route to constructing a bounce or rip is to assume a desired time dependence for the scale factor $a(t)$, or equivalently for $H(t)$, and then reconstruct the corresponding densities \cite{Myrzakulov:2014hva}; an application of this strategy to IDE is given in \cite{Brevik:2014lpa}. More recently, \cite{Burkmar:2025svw} showed that non-linear interactions combined with quadratic equations of state can yield singularity-free evolution from dynamical-system effects. The closest related work to ours is the interacting vacuum--matter scenario with linear couplings studied in \cite{Bruni:2021msx}, where the interaction generates an effective negative-energy contribution that can induce a bounce even in flat FLRW cosmologies. Our framework differs in that we assume a linear dark energy equation of state $w_{\rm de}=w$, whereas \cite{Burkmar:2025svw} employs a quadratic equation of state and \cite{Bruni:2021msx} assumes $w_{\rm de}=-1$. Moreover, the turnarounds we consider arise through explicit sign changes in the dark-sector densities driven by strong energy transfer.

Part of the aim of this study is to help fill this gap by providing explicit mechanisms through which crunching and bouncing scenarios can arise within IDE cosmology. It is well known that a non-singular bounce in general relativity typically requires either positive spatial curvature, a modification of the Friedmann equations, or an effective component that violates the Null Energy Condition (NEC), as in quintom-type matter models \cite{Cai:2007qw,Cai:2009zp}. Since we assume a flat FLRW universe within general relativity, any turnaround must be associated with NEC violation \cite{Hawking:1970zqf, Novello:2008ra}. In the IDE scenarios considered here, this occurs through negative dark-sector energy densities that arise in specific regions of parameter space, discussed further in Section \ref{sec:Conditions_general}. We emphasise that these regimes may suffer from gradient or ghost instabilities similar to those encountered in other NEC-violating bounce models \cite{Cline:2003gs,Vikman:2004dc}. Our goal is therefore not to construct realistic alternatives to standard cosmology, but to demonstrate that bounces, crunches, and cyclic behaviour can emerge naturally, together with their associated pathologies, within underexplored regions of IDE parameter space. These results motivate future work to determine whether any such scenarios yield testable predictions consistent with observations, or whether the relevant parameter regions should instead be excluded on physical grounds.

Our paper is structured as follows:
\begin{itemize}
    \item In Section \ref{sec:background}, we provide the background equations we use throughout the study and discuss the phenomenological interaction models that we consider. 
    
    \item In Section \ref{sec:neg}, we discuss how the negative energy densities required for bouncing and crunching cosmologies arise from dark sector interactions, as visualized in Figures \ref{fig:neg_de_visual} and \ref{fig:neg_dm_visual}, while phase portraits of the same mechanisms are presented in Appendix \ref{sec:DSA}. We also provide a brief overview of negative energies in other cosmological theories. 
    
    \item Section \ref{sec:Conditions_general} provides a mathematical outline of the required conditions for a turnaround associated with a crunch or bounce to occur, and the essential role of the SEC and NEC violation, as summarised in Table \ref{tab:bounce_flat_con}. 
    
    \item In Section \ref{sec:crunch} and \ref{sec:bounce}, we use both analytical solutions and dynamical system analysis techniques to describe the background dynamics of a flat universe with interacting dark sectors that allow for either a future big crunch (see Figure \ref{fig:crunch_5} and \ref{fig:DSA_H_Qdm}) or a past non-singular bounce (see Figure \ref{fig:bounce_5} and \ref{fig:DSA_H_Qde}), respectively. A combination of these scenarios to produce a cyclical model is realised in Appendix \ref{sec:cyclic}, and illustrated in Figure \ref{fig:cyclic_5}. 
    New analytical solutions for the minimum and maximum scale factor, and existence conditions for turnarounds for five special cases of the kernel $Q= 3 H (\delta_{\text{dm}} \rho_{\text{dm}} + \delta_{\text{de}}  \rho_{\text{de}})$ are derived in Appendix \ref{sec:mimmax}.

    \item In Section \ref{sec:otherQ}, we provide a brief note on the possibilities and difficulties of realising similar scenarios in other phenomenological IDE models. We also show in Figure \ref{fig:crunch_bounce_limit} and \ref{fig:DSA_H_Qdm+de} that both a bounce and a crunch can be realised for the same model, depending on the initial conditions. 
   
    \item Finally, we discuss our results and draw conclusions in Section \ref{sec:disc}.
     We also provide a summary of the updated parameter space for two IDE kernels in Table \ref{tab:IDE_parameter_space_summary}, which is also visualized in Figure \ref{fig:Q_parameter}.

\end{itemize}

\section{Background equations} \label{sec:background}
The background expansion of the universe is described by the Friedmann equations, which describe how the scale factor $a$ and its derivatives evolve over time:
\begin{gather} \label{F1}
\begin{split}
H^2 = \bigg( \frac{\dot{a}}{a} \bigg)^2 &= \frac{8}{3} \pi G \rho - \kappa \frac{c^2}{a^2},
\end{split}
\end{gather} 
\begin{gather} \label{F2}
\begin{split}
\bigg( \frac{\ddot{a}}{a}\bigg)  &= - \frac{4\pi G}{3}  \left(\rho + 3\frac{P}{c^2} \right),
\end{split}
\end{gather} 
where $\rho$ is the total energy density of all cosmological fluids and $P$ their total pressure, while $\kappa$ indicates the curvature of the universe. Equation \eqref{F2} describes the acceleration of the expansion and is also known as the Raychaudhuri equation. A useful way to relate the energy densities and pressures of all the different fluids is to introduce the total effective equation of state:
\begin{gather} \label{DSA.omega_eff_tot}
\begin{split}
w^{\rm{eff}}_{\rm{tot}} 
= \frac{P_{\rm{tot}}}{\rho_{\rm{tot}}} 
= \frac{w_{\rm{r}} \Omega_{\rm{r}} + w_{\rm{bm}} \Omega_{\rm{bm}} + w_{\rm{dm}} \Omega_{\rm{dm}} + w_{\rm{de}} \Omega_{\rm{de}}}
{\Omega_{\rm{r}} + \Omega_{\rm{bm}} + \Omega_{\rm{dm}} + \Omega_{\rm{de}}}.
\end{split}
\end{gather}
Expression \eqref{DSA.omega_eff_tot} is useful because $w^{\rm{eff}}_{\rm{tot}}$ determines whether the expansion or contraction is accelerating or decelerating.
\begin{gather} \label{DSA.omega_eff_tot_ACC}
\begin{split}
\text{If the universe is modelled as a single effective fluid   }  \; &\begin{cases}
w^{\rm{eff}}_{\rm{tot}} > -\frac{1}{3} \quad \rightarrow \quad \text{decelerating expansion,} \\
w^{\rm{eff}}_{\rm{tot}} < -\frac{1}{3} \quad \rightarrow \quad \text{accelerated expansion}. \\
\end{cases}
\end{split}
\end{gather}
It should be kept in mind that when the universe is contracting, then $w^{\rm{eff}}_{\rm{tot}} > -\frac{1}{3}$ implies that the contraction is happening quicker, which may lead to a big crunch. Conversely, $w^{\rm{eff}}_{\rm{tot}} < -\frac{1}{3}$ means that the rate of contraction is slowing down and a bounce may occur, followed by a period of expansion.

For this study, we will be concerned with models that allow energy transfer between the dark sectors. The simplest fluid models are usually described by coupled conservation equations of dark matter and dark energy, which conserve the total dark sector, but not the individual species, such that
\begin{gather} \label{eq:conservation.1}
\begin{split}
\dot{\rho}_{\text{dm}} + 3H \rho_{\text{dm}} = Q\;, \quad & \quad  \dot{\rho}_{\text{de}} + 3H (1 + w) \rho_{\text{de}} = -Q\;,\end{split}
\end{gather}
where $\rho_{\rm dm}$ and $\rho_{\rm de}$ are the energy densities of the dark matter and dark energy, $w$ is the equation of state for dark energy, $H$ is the Hubble parameter and $Q$ is a function that determines how energy is transferred between the dark fluids. In the simplest interaction models, $Q$ is phenomenological and proportional in a linear manner to one or both of the dark components. The interaction is normally also proportional to the Hubble parameter $H$ due to dimensional considerations, but this choice has been argued to be related to a possible temperature dependence of the interaction strength or other arguments from thermodynamics \cite{Valiviita:2008iv, Valentino_2020_DE, Nunes_2022}. Alternative models exist where the interaction rate is instead proportional to $H_0$ or a different constant \cite{Yang_2021_no_H}. Specifically, we will consider special cases of the general linear interaction kernel recently studied in \cite{vanderWesthuizen:2025I}:
\begin{gather} \label{eq:Q}
\begin{split}
Q=3H(\delta_{\rm dm}\rho_{\rm dm} + \delta_{\rm de}\rho_{\rm de}),
\end{split}
\end{gather}
where $\delta_{\rm dm}$ and $\delta_{\rm de}$ are dimensionless coupling constants which determine the dependency of the energy transfer rate to the dark matter and dark energy densities, respectively.
In our study, we will mostly look at the special cases where $\delta_{\rm dm}=0$ (in Section \ref{sec:crunch}) or $\delta_{\rm de}=0$ (in Section \ref{sec:bounce}),  while a brief discussion of the more general cases are provided in Section \ref{sec:otherQ} and Appendix \ref{sec:mimmax}. 

\section{Negative energy in cosmology} \label{sec:neg}
\subsection{Negative energies and instabilities in the strong coupling regime}

The negative energies required for a turnaround in a flat cosmological model in general relativity arise from the lack of an energy-transfer braking mechanism. Thus, we often find $Q\neq0$ when either $\rho_{\text{dm}}=0$ or $\rho_{\text{de}}=0$, causing the energy densities of one or both of the dark species to cross into negative values. For the interaction kernel \eqref{eq:Q}, in-depth discussion of how negative energies arise from both dynamical system analysis considerations and from the analytical solutions for $\rho_{\rm dm}$ and $\rho_{\rm de}$ are found in \cite{vanderWesthuizen:2023hcl, vanderWesthuizen:2025I}. For many fluid models, negative energy densities are inevitable when energy flows from dark matter to dark energy (the iDMDE regime with $Q<0$), corresponding to $\delta<0$ in an expanding universe. Conversely, a small energy transfer from dark matter to dark energy (iDEDM regime with $Q>0$ and $\delta>0$) causes the dark components to dilute at the same rate, achieving a positive constant ratio in the asymptotic past and future, thus solving the coincidence problem with positive energies at all times.

\begin{figure}[p]
    \centering

    \begin{minipage}{0.95\linewidth}

        \centering
        \includegraphics[width=0.9 \linewidth]{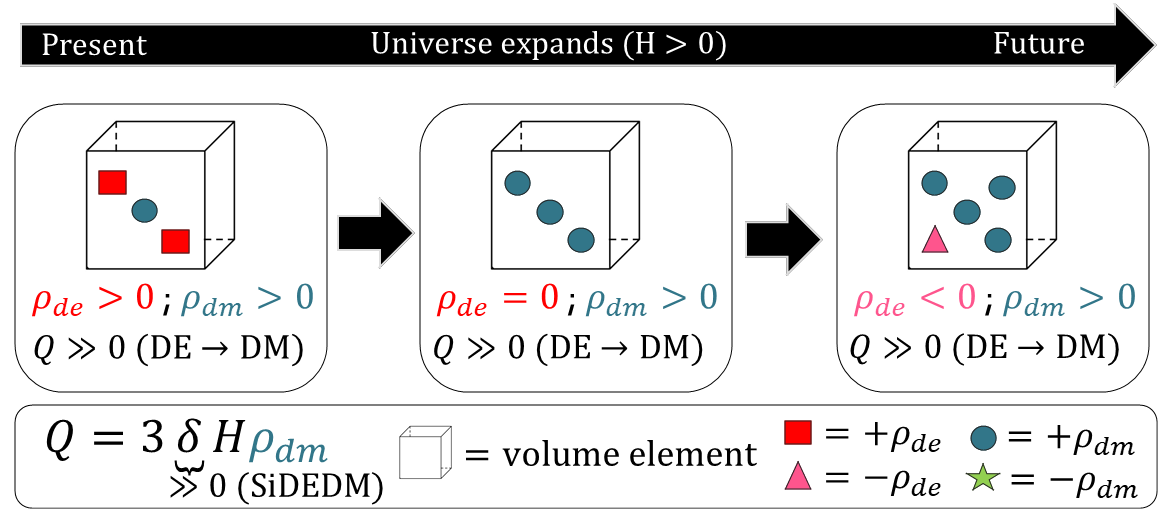}
        \caption{Visualization of how negative dark energy $\rho_{\rm de}<0$ may arise in the future for the interaction model $Q= 3 H \delta \rho_{\text{dm}}$, given the case where there is a strong coupling, $\delta\gg 0$, that causes energy transfer from dark energy to dark matter, here named the Strong interacting Dark Energy Dark Matter (SiDEDM) regime. For this specific coupling, the SiDEDM corresponds to the upper limit  $\delta>-\dfrac{w}{1+r_0}$ that causes negative energies, which has been previously derived from both dynamical system analysis considerations \cite{vanderWesthuizen:2025I} (illustrated here in Figure \ref{fig:2D_Q_general_Phase_portrait_boundaries}) and from the analytical solutions for the evolution of dark energy \cite{vanderWesthuizen:2023hcl}. We consider a volume element of space as we move from the present into the future, the universe expands $H>0$. The strong energy transfer causes dark energy to dilute and dark matter to increase, but we see that $Q\neq0$ when $\rho_{\rm de}=0$, thus energy transfer continues which causes negative dark energy to emerge. This feature may lead to future big crunch, as discussed in Section \ref{sec:crunch}.}
        \label{fig:neg_de_visual}
    \end{minipage}
    \vspace{1cm}

    \begin{minipage}{0.95\linewidth}

        \centering
        \includegraphics[width=0.9 \linewidth]{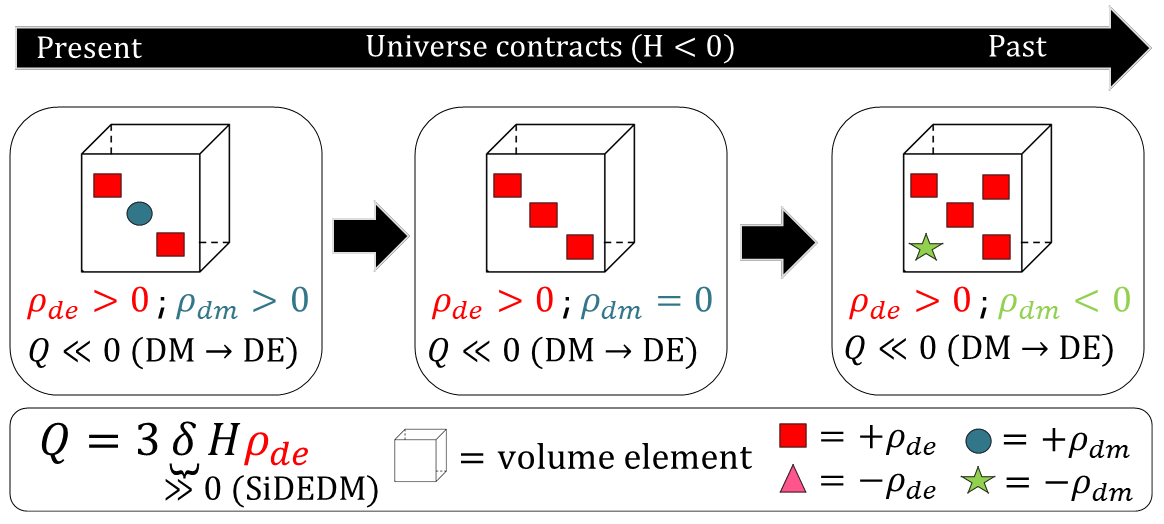}
        \caption{Visualization of how negative dark matter $\rho_{\rm dm}<0$ may arise in the past for the interaction model $Q= 3 H \delta \rho_{\text{de}}$, given the case where there is a strong coupling, $\delta\gg 0$, that causes energy transfer from dark energy to dark matter (SiDEDM regime). For this specific interaction, the SiDEDM corresponds to the upper limit  $\delta>-\dfrac{w}{1+1/r_0}$ which causes negative dark matter densities, which has been previously derived from both dynamical system analysis considerations \cite{vanderWesthuizen:2025I} (illustrated here in Figure \ref{fig:2D_Q_general_Phase_portrait_boundaries}) and from the analytical solutions for the evolution of dark matter \cite{vanderWesthuizen:2023hcl}. We consider a volume element of space as we move back in time from the present into the past such that the universe contracts $H<0$ and the energy transfer is from dark matter to dark energy. Dark energy increases and dark matter dilutes away, but since $Q\neq0$ when $\rho_{\rm dm}=0$, the energy transfer is not stopped when no dark matter remains, causing negative dark matter to emerge. This feature may lead to past non-singular bounce, as discussed in Section \ref{sec:bounce}.}
        \label{fig:neg_dm_visual}
    \end{minipage}

\end{figure}

We focus on the upper limits that were derived for $\delta\gg0$ in \cite{vanderWesthuizen:2023hcl, vanderWesthuizen:2025I, vanderWesthuizen:2025II, vanderWesthuizen:2025III}, corresponding to a strong coupling that causes significant energy transfer from dark energy to dark matter such that dark energy dilutes faster than dark matter, here named the Strong interacting Dark Energy Dark Matter (SiDEDM) regime. Unlike the iDEDM regime, which always has positive energies, the SiDEDM regime can lead to negative energies as well.  A brief discussion on how negative energies in the SiDEDM regime arise from dynamical systems considerations are given in Appendix \ref{sec:DSA}. 
For an intuitive understanding of how negative energies emerge in a given volume of space for two of the special cases of \eqref{eq:Q} in the SiDEDM regime, we have created the visualizations of the processes in Figure \ref{fig:neg_de_visual} and \ref{fig:neg_dm_visual}, with corresponding phase portraits in Figure \ref{fig:2D_Q_general_Phase_portrait_boundaries}, for $Q= 3 H \delta \rho_{\text{dm}}$ and $Q= 3 H \delta \rho_{\text{de}}$, respectively.
The different regimes considered in this paper are given in \eqref{Q_directions}, which only apply for the linear models of the form \eqref{eq:Q}.
\begin{equation}
Q = 
\begin{cases} 
< 0 & \text{Dark Matter} \rightarrow \text{Dark Energy (iDMDE regime, with $\rho_{\rm dm/de}<0$)},\\ 
= 0 & \text{No interaction (with $\rho_{\rm dm/de}>0$). }\\
> 0 & \text{Dark Energy} \rightarrow \text{Dark Matter (iDEDM regime, with $\rho_{\rm dm/de}>0$)},\\ 
\gg 0 & \text{Dark Energy} \rightarrow \text{Dark Matter (SiDEDM regime, with $\rho_{\rm dm/de}<0$)}.
\end{cases}
\label{Q_directions}
\end{equation}

We stress that negative energy densities appearing in the background evolution should be interpreted as effective fluid descriptions arising from the phenomenological interaction, rather than as fundamental violations of energy positivity at the microscopic level.

Besides negative energies, interacting dark energy models may exhibit instabilities at the level of linear perturbations, depending on the form and strength of the interaction. A commonly used diagnostic for such behaviour is the so-called doom factor, introduced in \cite{M.B.Gavela_2009} and applied in~\cite{Salvatelli:2013wra,Bachega:2019fki,DiValentino:2020vnx,Lucca:2021dxo,Yang:2021hxg,Forconi:2023hsj,Giare:2024ytc,Sabogal:2024yha}, which signals the presence of early-time non-adiabatic instabilities in the dark energy perturbations within the standard fluid description. Within this framework, it has been shown that regimes in which energy flows from dark energy to dark matter $Q>0$ and the dark energy equation of state satisfies $w<-1$ are typically free from such instabilities according to the doom-factor criterion. This suggests that parts of the strong-interaction regime considered here may evade early-time perturbative instabilities, although a dedicated perturbative analysis is required to assess stability in the specific scenarios studied in this work. Since the present paper is restricted to background dynamics, we do not perform such an analysis here. For those further interested in perturbations that include dark sector interactions, scalar perturbations are studied in \cite{Valiviita:2008iv, Valiviita:2009nu, M.B.Gavela_2009, Majerotto:2009np, Honorez_2010, Yang:2018euj, Johnson:2020gzn}, while new studies on linear and non-linear growth are found in \cite{Escobal:2026lnp} and \cite{Tudes:2024jpg, Silva:2025bnn}, respectively. We note that several approaches have been proposed to consistently extend perturbation studies beyond the standard fluid framework, including the parametrized post-Friedmann formalism \cite{Li_2014, Li_2014_2, Skordis_2015, Zhang_2017, Feng_2018, Dai_2019, Li_2020_2, Li_2023}, which can substantially enlarge the phenomenologically viable parameter space.

 \subsection{Precedent for negative dark energy and dark matter in other theories} \label{sec:neg_lit}
The possibility of negative energy density components may be deemed unphysical and seem to either immediately rule out the model, or at least the part of the parameter space where these negative energies exist. Interestingly, there is some precedent in the literature for negative dark energy or a negative cosmological constant (NGC). Another class of models exist where the dark energy switches sign at some point in cosmic evolution.

An example of a sign-switching cosmological constant is given by the $\Lambda_{s}$CDM model, where the universe switches from a negative anti-de Sitter (AdS) vacuum $\Lambda_{\rm{Ads}}<0$ at high redshift to a positive de Sitter (dS) vacuum  $\Lambda_{\rm{ds}}>0$ at some smaller redshift and into the present \cite{Akarsu:2019hmw, Akarsu:2021fol}. Comparisons with data provide preferences for models where the transition is quite rapid, which additionally avoids pathologies which may otherwise arise in astrophysical systems \cite{Akarsu:2022typ, Paraskevas:2023itu}. Furthermore, at the AdS-dS transition dark energy is equal to zero, which in turn causes the dark energy effective equation of state to diverge such that we have $w_{\rm{DE}}^{\rm{eff}}= \pm \infty$ at the transition \cite{Ozulker:2022slu, Adil:2023ara, Menci:2024rbq}. 

This switch from a negative dark energy component ($\rho_{\rm{de}}<0$ at large $z$) to a positive one ($\rho_{\rm{de}}>0$ at small $z$) near the present at $z_{\text{(de=0)}}$, which is associated with divergent phantom crossing for $w^{\rm{eff}}_{\rm{de}}$, is also predicted for the familiar interacting dark energy kernels $Q= 3 H (\delta_{\text{dm}} \rho_{\text{dm}} + \delta_{\text{de}}  \rho_{\text{de}})$, and its special cases  $Q=3H\delta( \rho_{\text{dm}}+\rho_{\text{de}})$, $Q=3H\delta( \rho_{\text{dm}}-\rho_{\text{de}})$ and $Q=3H\delta \rho_{\text{dm}}$, as well as the non-linear interaction $Q=3\delta H  \left(\frac{\rho_{\text{dm}}^2}{\rho_{\text{dm}}+\rho_{\text{de}}} \right)$, when energy flows from dark matter to dark energy, corresponding to $\delta_{\rm{dm}}<0$ and $\delta<0$, as shown in \cite{vanderWesthuizen:2025I, vanderWesthuizen:2025II}. A preference for this sign-switching behaviour in the phenomenological IDE models studied here were also found in \cite{Figueruelo2026IDEconstraints}.  It should be noted that unlike $\Lambda_{s}$CDM scenarios where $w(z)=-1$ as $z\rightarrow\infty$ and $z\rightarrow-1$, interaction models tend to have different asymptotic past and future behaviour, depending on the kernel and strength of the interaction.

Beyond energy exchange between the dark sectors that can produce sign-changing dark energy, other possible mechanisms for a AdS-dS transition can be found from modifications of gravity  \cite{Akarsu:2024nas, Souza_2024, Akarsu:2024qsi}, or string theory and extra dimension inspired approaches \cite{Anchordoqui:2024gfa, Anchordoqui:2023woo, Anchordoqui:2024dqc}. Specifically, within string theory it is difficult to construct a dS vacuum, which often leads to a preference for an AdS vacuum, corresponding to a negative dark energy component, which has been explored alongside the swampland conjecture, and symmetry considerations using the AdS/CFT correspondence \cite{OColgain:2018czj, Kinney:2018nny, Moritz:2017xto, Obied:2018sgi}. For recent observational constraints on models with sign-switching AdS-to-dS behaviour or a negative cosmological constant, see \cite{Pedrotti:2025ccw}. An AdS vacuum or negative cosmological constant can exist alongside a positive dark energy component, whose behaviour has recently been reconstructed using DESI BAO data \cite{Wang:2026kbg}.  Phantom crossings can also be enabled with a negative quintessence dark energy component, as shown in \cite{Gomez-Valent:2025mfl, Gonzalez-Fuentes:2025lei}. For further discussions, observational constraints, and reconstructions of models featuring negative dark energy densities, see \cite{V_liviita_2010, BOSS:2014hwf, Poulin:2018zxs, Wang:2018fng, Visinelli:2019qqu, Calderon:2020hoc, Ong:2022wrs, Malekjani:2023ple, Bouhmadi-Lopez:2025spo, Menci:2026ajy, Akarsu:2026anp, Ibarra-Uriondo:2026zbp, Bouhmadi-Lopez:2026ckz, Gokcen:2026pkq}. 

Conversely, negative dark matter has less often been invoked, but so-negative mass cosmology (NMC) models suggesting negative mass for the dark components include Bondi negative-mass constructions \cite{Bondi:1957zz, Najera:2021tcx, Manfredi:2026qoz}, Dirac-milne models where anti-matter is given a negative active gravitational mass \cite{Benoit-Levy:2011esx} or Farnes-type models where dark phenomena can be unified into a single negative mass fluid \cite{Farnes:2017gbf}. These negative mass components are often invoked to drive cosmic acceleration, as can be seen from the Raychaudhuri equation applied to negative mass components \eqref{eq:dm_neg_repulsive}. Refutations of the validity of Farnes-type models and negative mass in general relativity are found in \cite{Socas-Navarro:2019pps}. A recent discussion on the possibility of negative mass objects (NMO) and their observational viability is found in \cite{Nojiri:2026ubx}.

\section{Conditions for Big Crunch and Big Bounce Cosmologies} \label{sec:Conditions_general}
In order for a cosmological model which is at present expanding to have either a big crunch in the future, or a non-singular bounce in the past, it is required that there should at some point be a transition for expansion to contraction, or vice versa. This transition, or turnaround of the scale factor, can only occur in flat FLRW models within general relativity given certain criteria, which violate the energy conditions described below. 

\subsection{Violation of Strong Energy Condition}  {\label{SEC}}
In bouncing scenarios, a past singularity is replaced by a smooth transition from contraction to expansion. To achieve this, the acceleration of the expansion of the universe must be positive. The acceleration is described by the Raychaudhuri equation \eqref{F2}, which we require to be positive, such that:\\
\begin{gather} \label{1.3.1}
\begin{split}
\text{Bounce Condition}: \underbrace{\left( \frac{\ddot{a}}{a}\right)}_{>0}  &= - \frac{4\pi G}{3} \underbrace{\left( \rho + 3\frac{P}{c^2} \right)}_{<0} \quad \rightarrow  \quad  \underbrace{\rho + 3P  <0}_{\text{Violates SEC}},\\
\text{Big Crunch Condition}: \underbrace{\left( \frac{\ddot{a}}{a}\right)}_{<0}  &= - \frac{4\pi G}{3} \underbrace{\left( \rho + 3\frac{P}{c^2} \right)}_{>0} \quad \rightarrow  \quad  \underbrace{\rho + 3P  >0}_{\text{Obeys SEC}},\\
\end{split}
\end{gather} \\
where we used natural units $c = 8 \pi G = 1$ to obtain the last equality (natural units are only used when stated explicitly). The equality in \eqref{1.3.1} for a single fluid is equivalent to $w<-\frac{1}{3}$ and violates the strong energy condition (SEC), which states that $\rho + 3P  	\geq 0$. One consequence of SEC is that the active gravitational potential is always positive \citep{Carloni:2005ii}. Violation of SEC is standard in cosmology since the observed accelerated expansion of the universe and the reintroduction of the cosmological constant which has $w=-1$.

\subsection{Violation of Null Energy Condition} {\label{NEC}}

As mentioned above, the universe must change from a period of contraction to a period of expansion in a bounce, and vice versa for a crunch. Contraction corresponds to a negative value of the Hubble parameter $(H<0)$, while expansion has a positive Hubble parameter $(H>0)$. Thus, for a bounce to occur, the time derivative of the Hubble parameter $\dot{H}$ must be positive for the Hubble parameter to change from negative to positive, and vice versa for a crunch, such that:
\begin{gather} \nonumber
\begin{split}
\text{Bounce Condition}: \quad \text{Contraction } (H<0) \rightarrow  \text{Bounce } (H=0, \;\dot{H}>0) \rightarrow\text{Expansion } (H>0), \\
\text{Big Crunch Condition}: \quad \text{Expansion } (H>0) \rightarrow  \text{Bounce } (H=0, \;\dot{H}<0) \rightarrow  \text{Contraction } (H<0).\\
\end{split}
\end{gather} \\
Where $\dot{H}$ can be shown to be equal to:
\begin{gather} \label{BCdH.1}
\begin{split}
\dot{H}  &=  \left(\frac{\ddot{a}}{a}\right) - H^2 =\frac{\kappa}{a^2} - \frac{1}{2} \left(\rho + P \right).  \\
\end{split}
\end{gather} \\
Here we note that the first term on the R.H.S. is the Raychaudhuri equation from \eqref{F2} and the second term is the Friedmann equation from \eqref{F1}. To obtain the final equality, we  substituted \eqref{F2} and \eqref{F1} into (\ref{BCdH.1}) and used natural units. 
Thus, if we consider the case of a flat universe $\kappa = 0$, the first terms disappear and since we need to obey the condition that $\dot{H} > 0$, we obtain that: 
 \begin{gather} \label{BCdH.3} 
\begin{split}
\text{Bounce Condition}:\underbrace{\dot{H}}_{>0}  &= - \frac{1}{2} \underbrace{\left(\rho + P \right)}_{<0} \quad \rightarrow  \quad  \underbrace{\rho + P  <0}_{\text{Violates NEC}},  \\
\text{Big Crunch Condition}:\underbrace{\dot{H}}_{<0}  &= - \frac{1}{2} \underbrace{\left(\rho + P \right)}_{>0} \quad \rightarrow  \quad  \underbrace{\rho + P  >0}_{\text{Obeys NEC}},  \\
\end{split}
\end{gather} \\
which for the bouncing case in a single fluid description is equivalent to $w<-1.$ This violates the Null Energy Condition (NEC) which states that inertial mass density is always positive: $\rho + P  \geq 0 $. A consequence of violating NEC is that fields may appear that have negative kinetic energy. These fields are known as ghosts and would preferably be avoided. Recent BAO observations have hinted at the possibility of $w<-1$ in the past \cite{DESI:2025fii}, with a recent discussion on this suggested possibility of NEC violation found in \cite{Caldwell:2025inn}. 
For other discussions on violations of energy conditions, see \cite{Carroll_2003, Santos_2007, Rubakov_2014, martinmoruno2017classicalsemiclassicalenergyconditions}.  \\ 
For both bouncing and crunching cosmologies, we require $H=0$ at the turnaround, and from the 1st Friedmann equation \eqref{F1}, it follows that we require:
\begin{gather} \label{BCH.1}
\begin{split}
\frac{8}{3} \pi G \rho &=  \frac{\kappa c^2}{a^2}  \quad \rightarrow \quad\rho_{\rm{bounce/crunch}}=  \frac{3\kappa c^2}{8 \pi Ga_{\rm{min/max}}^2}.\\
\end{split}
\end{gather} \\
For condition \eqref{BCH.1} to hold alongside $\rho>0$, we require a universe with closed curvature $k=+1$ or $\Omega_{\rm{k}}<0$. Conversely, for a flat universe ($k=+0$) to obey both \eqref{BCdH.3} and \eqref{BCH.1},  we require that:
\begin{gather} \label{BCH.2}
\begin{split}
\text{Flat Universe  Bounce/Crunch Condition : }\rho_{\rm{tot}}=0 \text{ and } P_{\rm{tot}}\neq0 \text{ at turnaround } (a_{\rm{min/max}}). \\
\end{split}
\end{gather} 
Condition \eqref{BCH.2} cannot hold in standard cosmology where all energy densities are positive at all times, but this condition can be met in IDE cosmology, where negative energy densities are a well-known feature of certain section of the parameter space \cite{vanderWesthuizen:2025I, vanderWesthuizen:2025II, vanderWesthuizen:2025III}. While individual interacting components may cross zero, the total energy density of the universe can never become negative $\rho_{\rm tot}<0$, as this would lead to imaginary values of $H$. It should be noted that the lower energy density limit $\rho_{\rm{tot}}=0$ does not imply $P=0$, since the individual components each still has their corresponding pressure even though the sum of their energy densities cancel out. Since we require $\rho_{\rm tot}=0$ at $H=0$, and $P_{\rm tot}=P_{\rm dm}+P_{\rm de}=w \rho_{\rm de}$, we can see from \eqref{BCdH.3} that the sign of the dark energy density will help us distinguish if the turnaround corresponds to a crunch or bounce: 
 \begin{gather} \label{BCdH.5} 
\begin{split}
\text{Bounce Condition}:\underbrace{\dot{H}}_{>0}  &=  \underbrace{-\frac{1}{2} w}_{>0} \underbrace{ \rho_{\rm de} }_{>0} \quad \rightarrow  \quad  \rho_{\rm de}>0 \;; \; \rho_{\rm dm}<0 \quad \text{for IDE models in a flat FLRW universe},  \\
\text{Big Crunch Condition}:\underbrace{\dot{H}}_{<0}  &=   \underbrace{-\frac{1}{2} w}_{>0} \underbrace{ \rho_{\rm de} }_{>0} \quad \rightarrow  \quad  \rho_{\rm de}<0 \;; \; \rho_{\rm dm}>0 \quad \text{for IDE models in a flat FLRW universe}.  \\
\end{split}
\end{gather} \\
The last equalities are obtained since we need $\rho_{\rm dm}$ and $\rho_{\rm de}$ to have opposite signs for $\rho_{\rm tot}=0$ at the turnaround. Finally, we emphasise that the NEC violation derived here is specific to flat FLRW cosmologies. In the presence of positive spatial curvature, the curvature term in $\dot{H}$ can drive a bounce even when all energies are positive, as seen in Appendix \ref{sec:cyclic} and Figure \ref{fig:bounce_5}.

\subsection{Summary of bounce and crunch conditions} \label{con}
The conditions required moments before, during and after the turnaround needed for a bounce or crunch in a flat FLRW universe with dark sector interaction in general relativity is given in Table \ref{tab:bounce_flat_con} and \ref{tab:crunch_flat_con}. 
\begin{table}[H]
\centering
\begin{tabular}{|c|c|c|c|c|c|c|c|}
\hline 
Phase & $a$ &  $H$ &  $\rho_{\rm{tot}}$ & $\rho_{\rm{dm}}$ & $\rho_{\rm{de}}$ &  SEC ($\rho_{\rm{tot}}+3P_{\rm{tot}}>0$ or $w^{\rm{eff}}_{\rm{tot}}>-\frac{1}{3}$)  &  NEC ($\rho_{\rm{tot}}+P_{\rm{tot}}>0$ or $w^{\rm{eff}}_{\rm{tot}}>-1$)   \\ 
\hline 
\hline 
Expansion & $+$  & $+$ & $+$ & $+$ & $-$   & Holds (Deceleration) & Holds   \\ 
\hline 
Turnaround & $a_{\rm{max}}$  & $0$ & $0$ & $+$ & $-$ & Holds (Deceleration) & Holds  \\ 
\hline 
Contraction & $+$  & $-$ & $+$  & $+$ & $-$ & Holds (Deceleration) & Holds  \\   
\hline 
\end{tabular}
\caption{Conditions required in a small time interval before, during and after a turnaround preceding a big crunch in a flat FLRW universe described by general relativity, allowing for dark sector interactions.}
\label{tab:crunch_flat_con}
\end{table}

\begin{table}[H]
\centering
\begin{tabular}{|c|c|c|c|c|c|c|c|}
\hline 
Phase & $a$ &  $H$ &  $\rho_{\rm{tot}}$ & $\rho_{\rm{dm}}$ & $\rho_{\rm{de}}$ &  SEC ($\rho_{\rm{tot}}+3P_{\rm{tot}}>0$ or $w^{\rm{eff}}_{\rm{tot}}>-\frac{1}{3}$)  &  NEC ($\rho_{\rm{tot}}+P_{\rm{tot}}>0$ or $w^{\rm{eff}}_{\rm{tot}}>-1$)     \\ 
\hline 
\hline 
Contraction & $+$  & $-$ & $+$ & $-$ & $+$ & Violated (Acceleration) & Violated   \\   
\hline 
Bounce & $a_{\rm{min}}$  & $0$ & $0$ & $-$ & $+$  &  Violated (Acceleration) & Violated   \\ 
\hline 
Expansion & $+$  & $+$ & $+$ & $-$ & $+$  & Violated (Acceleration) & Violated   \\ 
\hline 
\end{tabular}
\caption{Conditions required in a small time interval before, during and after a turnaround preceding a big bounce in a flat FLRW universe described by general relativity, allowing for dark sector interactions.}
\label{tab:bounce_flat_con}
\end{table}

\section{Future Big Crunch case study: $(Q= 3 H \delta \rho_{\text{dm}})$}  \label{sec:crunch}
\subsection{Mathematical background}
We have observed that the turnaround preceding a big crunch will occur when $\rho_{\rm{tot}}=0$. In this model, we assume the future will be dominated by dark matter and dark energy, so for \eqref{BCH.2} to hold, we need \eqref{BC.1} to hold.
\begin{gather} \label{BC.1}
\begin{split}
\rho_{\rm{tot}}= \rho_{\rm{dm}}+\rho_{\rm{de}}&= 0 \text{ at turnaround } (a_{\rm{min/max}}) \\
\rightarrow \rho_{\rm{dm}}&=-\rho_{\rm{de}}.
\end{split}
\end{gather} 

\begin{figure}
    \centering
    \includegraphics[width=0.9\linewidth]{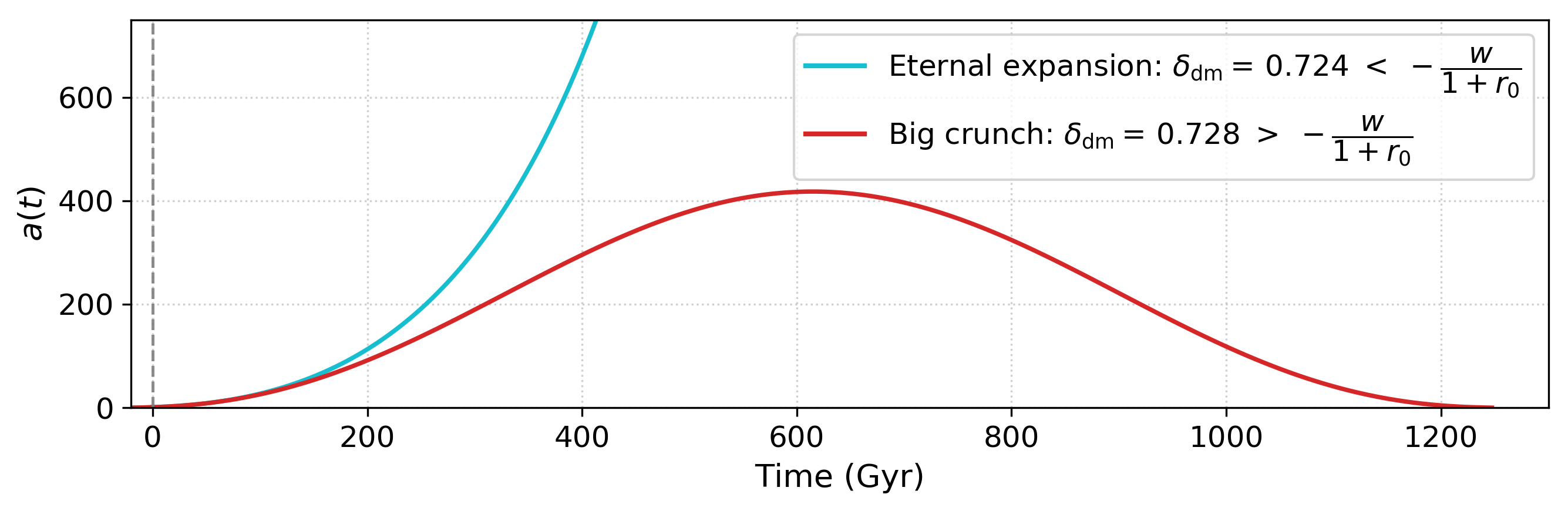}
    \caption{Evolution of scale factor $a$ vs time for interacting dark energy model $Q= 3 H \delta \rho_{\text{dm}}$. For both cases we have set $\Omega_{\rm{(bm,0)}}=0.05$, $\Omega_{\rm{(dm,0)}}=0.26$, $\Omega_{\rm{(de,0)}}=0.69$, $w=-1$. From these parameters, the upper positive energy limit is $\delta = -\frac{w }{1+r_0}=0.726$. If the interaction strength is below this limit ($\delta =0.724< -\frac{w }{1+r_0}$) the universe expands forever, while if above the limit ($\delta=0.728 > -\frac{w }{1+r_0}$), negative dark energy emerges and the universe eventually re-collapses in the future. A wide range of initial conditions that allow for negative dark energy can cause a big crunch turnaround, as seen in the phase portraits in Figure \ref{fig:DSA_H_Qdm}. The negative energy densities emerge from the mechanisms illustrated in Figure \ref{fig:neg_de_visual} and \ref{fig:2D_Q_general_Phase_portrait_boundaries}.}
    \label{fig:crunch_limit}
\end{figure}

For the special case where $\delta_{\rm de}=0$, we have the interaction kernel $Q= 3 H \delta \rho_{\text{dm}}$, for which the dark matter and dark energy evolves with the following equations \cite{vanderWesthuizen:2025I}:
\begin{gather}
\begin{split} \label{eq:rho_dm_Q_dm}
\rho_{\text{dm}} = \rho_{\text{(dm,0)}} a^{-3(1 -\delta)},
\end{split}   
\end{gather}
\begin{gather}
\begin{split} \label{eq:rho_de_Q_dm}
\rho_{\text{de}} = \left(\rho_{\text{(de,0)}} + \rho_{\text{(dm,0)}} \left( \frac{\delta}{\delta + w} \right) \left[  1 - a^{3 (\delta+ w)} \right] \right) a^{-3(w + 1)}.
\end{split}   
\end{gather}
We can see that $\rho_{\rm dm}$ will always be positive, while $\rho_{\rm de}$ may become negative, so from the conditions in \eqref{BCdH.5} a turnaround will be associated with a crunch for this model, as it cannot experience a bounce. The positive energy conditions for this model requires energy flow from DE to DM with the coupling constant in the range $0\le\delta \le -\frac{w }{1+r_0}$. This upper limit is associated with negative DE in the future \cite{vanderWesthuizen:2023hcl}, which has an attractive effect that can cause an eventual re-collapse of the universe. This attractive effect of negative DE arises from the combined effects of the energy density and pressure in the Raychaudhuri equation \eqref{F2} (in natural units), with $p_{\rm{de}}=w\rho_{\rm{de}}$, as shown in \eqref{eq:de_neg_attractive}.
\begin{equation}
\begin{split}
\text{Negative DE is attractive}: \underbrace{\bigg( \frac{\ddot{a}}{a}\bigg)}_{<0}  &=\underbrace{ - \frac{4\pi G}{3}  \left[ \underbrace{\rho_{\rm{de}}}_{<0} \underbrace{(1+  3w_{\rm{de}})}_{<0} \right]}_{<0}.
\end{split}
\label{eq:de_neg_attractive}
\end{equation}
We find that the exact condition for a re-collapse is \eqref{eq:Q_dm_crunch_con}, derived in Appendix \ref{sec:mimmax} from the requirement of a positive and real solution for $a_{\rm min/max}$ in \eqref{eq:Q_dm_a_max}. This condition is the upper positive energy bound in for energy transfer from dark energy to dark matter derived in \cite{vanderWesthuizen:2023hcl, vanderWesthuizen:2025I}, as shown in Figure \ref{fig:crunch_limit} (for all figures, the input parameters are $H_0=67.4$ km/s/Mpc, $\Omega_{\rm{(r,0)}}=9\times10^{-5}$, $\Omega_{\rm{(bm,0)}}=0.049$, $\Omega_{\rm{(dm,0)}}=0.266$, $\Omega_{\rm{(de,0)}}=0.685$, $w=-1$, with only $\delta$ varying between figures, unless stated otherwise). 
\begin{equation}
\begin{split}
\text{Flat Universe Big Crunch Condition : }\quad \delta > -\frac{w }{1+r_0}  \quad \text{(SiDEDM regime)}. 
\end{split}
\label{eq:Q_dm_crunch_con}
\end{equation}
where the present ratio of the dark components is $r_0=\rho_{(\rm dm,0)}/\rho_{(\rm de,0)}=\Omega_{(\rm dm,0)}/\Omega_{(\rm de,0)}$.

A turnaround from expansion to contraction will occur when the universe reaches its maximum size $a_{\rm{max}}$ when $\rho_{\rm{dm}}\approx-\rho_{\rm{de}}$. An expression for $a_{\rm{max}}$ is obtained by substituting \eqref{eq:rho_dm_Q_dm} and \eqref{eq:rho_de_Q_dm} into \eqref{BC.1} and solving for $a$, yielding:
\begin{equation}
\begin{split}
a_{\rm{max}}  \approx& \left[- \frac{\Omega_{\text{(de,0)}}}{\Omega_{\text{(dm,0)}}} \left(\frac{\delta+w}{w} \right) -\frac{\delta}{w}   \right]^{\frac{1}{{3(\delta +w) }}}. \\
\end{split}
\label{eq:Q_dm_a_max}
\end{equation}
We can see that this approximation holds when we substitute in the parameters $\Omega_{\rm{(dm,0)}}=0.26$, $\Omega_{\rm{(de,0)}}=0.69$, $w=-1$, and $\delta=+0.9$ into \eqref{eq:Q_dm_a_max}, which given $a_{\rm{max}}=4.55$. This matches the values for $a_{\rm{max}}$ seen in the top panel of Figure \ref{fig:crunch_5}, which is created with the same parameters.

\subsection{Step-by-step big crunch mechanism} \label{sec:crunch_steps}

\begin{figure}
    \centering
    \includegraphics[width=0.9\linewidth]{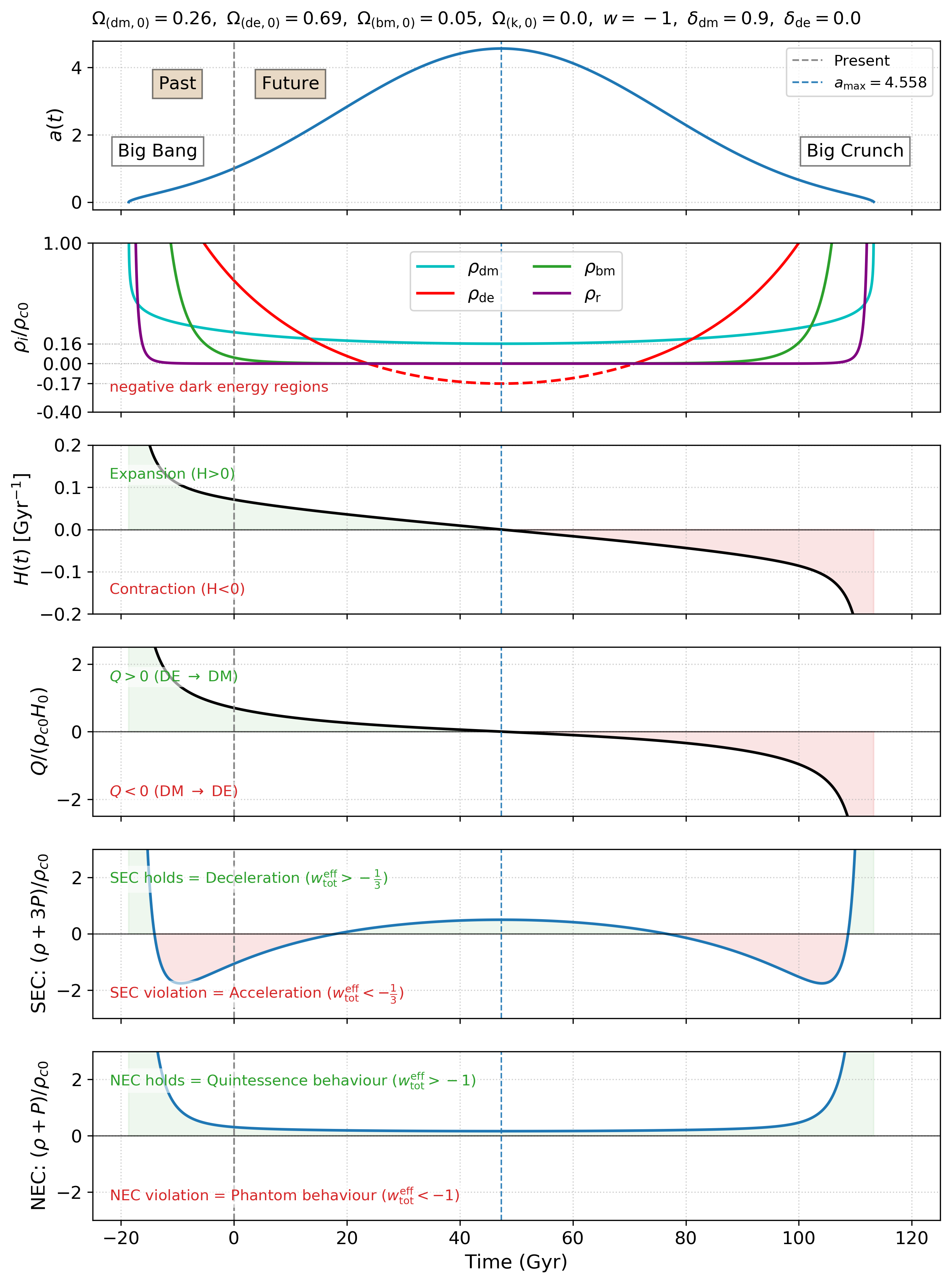}
    \caption{ Evolution of the scale factor 
    $a(t)$, dimensionless energy densities $\rho_i/\rho_{c0}$, Hubble parameter $H(t)$, dimensionless interaction term $Q/(\rho_{c0}H_0)$, and the strong and null energy conditions (SEC and NEC) for interacting dark energy model $Q= 3 H \delta \rho_{\text{dm}}$. As illustrated in Figure \ref{fig:neg_de_visual}, as energy is transferred from DE to DM, at some point DE becomes negative since $Q>0$ when $\rho_{\rm{de}}=0$, and the universe starts to decelerate $w^{\rm{eff}}_{\rm{tot}}<-\frac{1}{3}$. After some time we find $\rho_{\rm{dm}}=-\rho_{\rm{de}}$, corresponding to $\rho_{\rm{tot}}=0$, $H=0$ (as well as $Q=0$), where the universe experiences a turnaround from expansion $H>0$ to contraction $H<0$. During contraction, the negative $H$ causes $Q<0$ and energy now transfers back from DM to DE, causing DE to become positive at some point and even causing acceleration again $w^{\rm{eff}}_{\rm{tot}}<-\frac{1}{3}$. The matter components dominate again as $a$ becomes small causing deceleration and faster contraction, and a future big crunch occurs where $a\rightarrow 0$ and $\rho_{\rm{tot}}\rightarrow \infty$. See Figure \ref{fig:DSA_H_Qdm} for phase portraits showing similar behaviour for a larger selection of initial conditions.}
    \label{fig:crunch_5}
\end{figure}

In this section, we consider a more detailed analysis of the same interaction kernel, with similar initial conditions to $\Lambda$CDM ($H_0=67.4$ km/s/Mpc, $\Omega_{\rm{(r,0)}}=9\times10^{-5}$, $\Omega_{\rm{(bm,0)}}=0.05$, $\Omega_{\rm{(dm,0)}}=0.26$, $\Omega_{\rm{(de,0)}}=0.69$, $w=-1$), except for the energy transfer ($\delta_{\rm dm}=0.9$) from dark energy to dark matter required by \eqref{eq:Q_dm_crunch_con} for a future big crunch. These input parameters predict a qualitatively familiar past expansion history which includes a past Big Bang singularity, followed by an initial period of decelerating expansion and a present era of accelerated expansion, but with a different predicted future expansion.
To understand intuitively how our IDE model leads to a future big crunch, we refer to Figure \ref{fig:crunch_5}, where we have plotted the evolution of the scale factor $a$, energy densities $\rho$ all components, the Hubble parameter $H$, the energy transfer direction $Q$, and the validity of the strong energy condition (SEC) and Null energy condition (NEC) vs time for interacting dark energy model $Q= 3 H \delta \rho_{\text{dm}}$. For ease of illustration, we have plotted many of the panels in Figure \ref{fig:crunch_5} with dimensionless units by dividing by the present critical density $\rho_{c0}$ and the present Hubble constant $H_0$, as needed from dimensional considerations. We will analyse Figure \ref{fig:crunch_5}, starting from the present and moving further into the future, mentioning important transitions along the way. The plots are made using the two Friedmann equations \eqref{F1} and \eqref{F2}, the evolution of the energy densities \eqref{eq:rho_dm_Q_dm} and \eqref{eq:rho_de_Q_dm}, and the energy conditions \eqref{1.3.1} and \eqref{BCdH.3}. \\ \\

\begin{enumerate}
    \item \underline{Present accelerating expansion ($a=1 ;\;\rho_{\rm de}>0;\; H>0 ; \; Q>0 \;(\text{DE}\rightarrow\text{DM}); \; \text{SEC violated; NEC holds}$)}: \\ \\
    The model starts with initial conditions similar to that of the flat $\Lambda$CDM model, with all energies positive, the NEC holding and the SEC violated to give accelerated expansion. The only modification is a strong coupling $Q\gg0$ that causes energy transfer from DE to DM, with $\delta$ exceeding the limit derived in \eqref{eq:Q_dm_crunch_con}.  
    \item \underline{Negative DE decelerating expansion ($1<a<a_{\rm{max}} ;\;\rho_{\rm de}<0;\; H>0 ; \; Q>0 \;(\text{DE}\rightarrow\text{DM}); \; \text{SEC \& NEC holds} $)}: \\ \\
    The large energy transfer from DE to DM eventually causes the accelerating power of DE to diminish until  $w^{\rm{eff}}_{\rm{tot}}>-\frac{1}{3}$, implying that the SEC holds and that expansion rate is decelerating. The energy transfer continues and DE eventually becomes negative ($\rho_{\rm de}<0$) and attractive as shown in \eqref{eq:de_neg_attractive}, further decelerating the expansion rate.
    \item \underline{Turnaround at maximum scale factor ($a=a_{\rm{max}} ;\;\rho_{\rm de}<0;\; H=0 ; \; Q=0 \;(\text{no transfer}); \; \text{SEC \& NEC holds} $)}: \\ \\
    The dark energy component eventually becomes negative enough so that $\rho_{\rm tot}=0$ as given in condition \eqref{BC.1}. From the first Friedmann equation \eqref{F1}, this leads to $H=0$ and the expansion to stop at a maximum scale factor determined by \eqref{eq:Q_dm_a_max}. Importantly, from the Raychaudhuri equation \eqref{F2}, $\dot{H}<0$, as shown by the SEC that still holds, which implies deceleration throughout the turnaround which will therefore be followed by a period of contraction. It should be noted that the total fluid effective equation of state description breaks down at the turnaround, as  $\rho_{\rm tot}=0$, which leads to $w^{\rm eff}_{\rm tot}=\frac{P_{\rm{tot}}}{\rho_{\rm{tot}}} =\infty$.
    \item \underline{Negative DE decelerating contraction ($a<a_{\rm{max}} ;\;\rho_{\rm de}<0;\; H<0 ; \; Q<0 \;(\text{DM}\rightarrow\text{DE}); \; \text{SEC \& NEC holds} $)}: \\ \\
    As the universe starts contracting, the direction of energy transfer switches (since $H<0$ and $Q \propto H$) and energy flows from DM to DE. This energy transfer continues and eventually the DE component becomes positive again, eventually causing an end to the decelerated contraction.
    
    \item \underline{Positive DE accelerating contraction ($a>0 ;\;\rho_{\rm de}>0;\; H<0 ; \; Q<0 \;(\text{DM}\rightarrow\text{DE}); \; \text{SEC violated; NEC holds}$)}: \\ \\
    In this phase, DE becomes positive enough to cause a slow down of the rate of contraction A$w^{\rm{eff}}_{\rm{tot}}<-\frac{1}{3}$ and for SEC to he violated again. In this specific case, the contraction is well under way and eventually the matter components dominate again. 
  
    \item \underline{Contraction to big crunch ($a\rightarrow0 ;\;\rho_{\rm de} \rightarrow\infty;\; H \rightarrow-\infty ; \; Q<0 \;(\text{DM}\rightarrow\text{DE}); \; \text{SEC \& NEC holds} $)}: \\ \\
    In the final phase, the matter domination leads to a final phase of faster contraction which ends in a big crunch singularity with infinite energy densities at $t_{\rm{crunch}}$, where $a\rightarrow0$, $\rho_{\rm de} \rightarrow \infty$ and $H\rightarrow -\infty$.
\end{enumerate}

It should be mentioned that the steps given above are just an example of how a big crunch can proceed, but some steps might be avoided given different initial conditions. In all cases, the steps corresponding to those in Table \ref{tab:crunch_flat_con} will be necessary. The main features of the IDE bounce shown in Figure \ref{fig:crunch_5} is summarised in Table \ref{tab:crunch_phases_ide}.

\begin{table}[H]
\centering
\renewcommand{\arraystretch}{1.25}
\begin{tabular}{|c|c|c|c|c|c|c|}
\hline
\textbf{Phase} 
& $\boldsymbol{a}$ 
& $\boldsymbol{H}$ 
& $\boldsymbol{\rho_{\rm de}}$ 
& $\boldsymbol{Q}$ 
& \textbf{SEC} 
& \textbf{NEC} \\
\hline
\hline
Present accelerating expansion 
& $1$ 
& $+$ 
& $+$ 
& $>0$ (DE$\rightarrow$DM) 
& Violated ($w^{\rm eff}_{\rm tot}<-\frac{1}{3}$) 
& Holds \\ 
\hline
Negative DE decelerating expansion 
& $1<a<a_{\rm max}$ 
& $+$ 
& $-$ 
& $>0$ (DE$\rightarrow$DM) 
& Holds ($w^{\rm eff}_{\rm tot}>-\frac{1}{3}$) 
& Holds \\ 
\hline
Turnaround (maximum scale factor) 
& $a_{\rm max}$ 
& $0$ 
& $-$ 
& $0$ (no transfer) 
& Holds ($\dot{H}<0$)  
& Holds \\ 
\hline
Negative DE decelerating contraction 
& $1<a<a_{\rm max}$ 
& $-$ 
& $-$ 
& $<0$ (DM$\rightarrow$DE) 
& Holds ($w^{\rm eff}_{\rm tot}>-\frac{1}{3}$) 
& Holds \\ 
\hline
Positive DE accelerating contraction 
& $0<a<1$ 
& $-$ 
& $+$ 
& $<0$ (DM$\rightarrow$DE) 
& Violated ($w^{\rm eff}_{\rm tot}<-\frac{1}{3}$) 
& Holds \\ 
\hline
Contraction to big crunch 
& $a\rightarrow0$ 
& $H\rightarrow-\infty$ 
& $\rho_{\rm de}\rightarrow\infty$ 
& $<0$ (DM$\rightarrow$DE) 
& Holds ($w^{\rm eff}_{\rm tot}>-\frac{1}{3}$)
& Holds \\ 
\hline
\end{tabular}
\caption{Sequence of cosmological phases leading from late--time acceleration to a big--crunch singularity in the interacting dark energy model. The sign of the interaction term $Q$ indicates the direction of energy transfer between dark energy (DE) and dark matter (DM).}
\label{tab:crunch_phases_ide}
\end{table}

\subsection{Crunch mechanism from phase portraits}
\begin{figure}
    \centering
    \includegraphics[width=0.95 \linewidth]{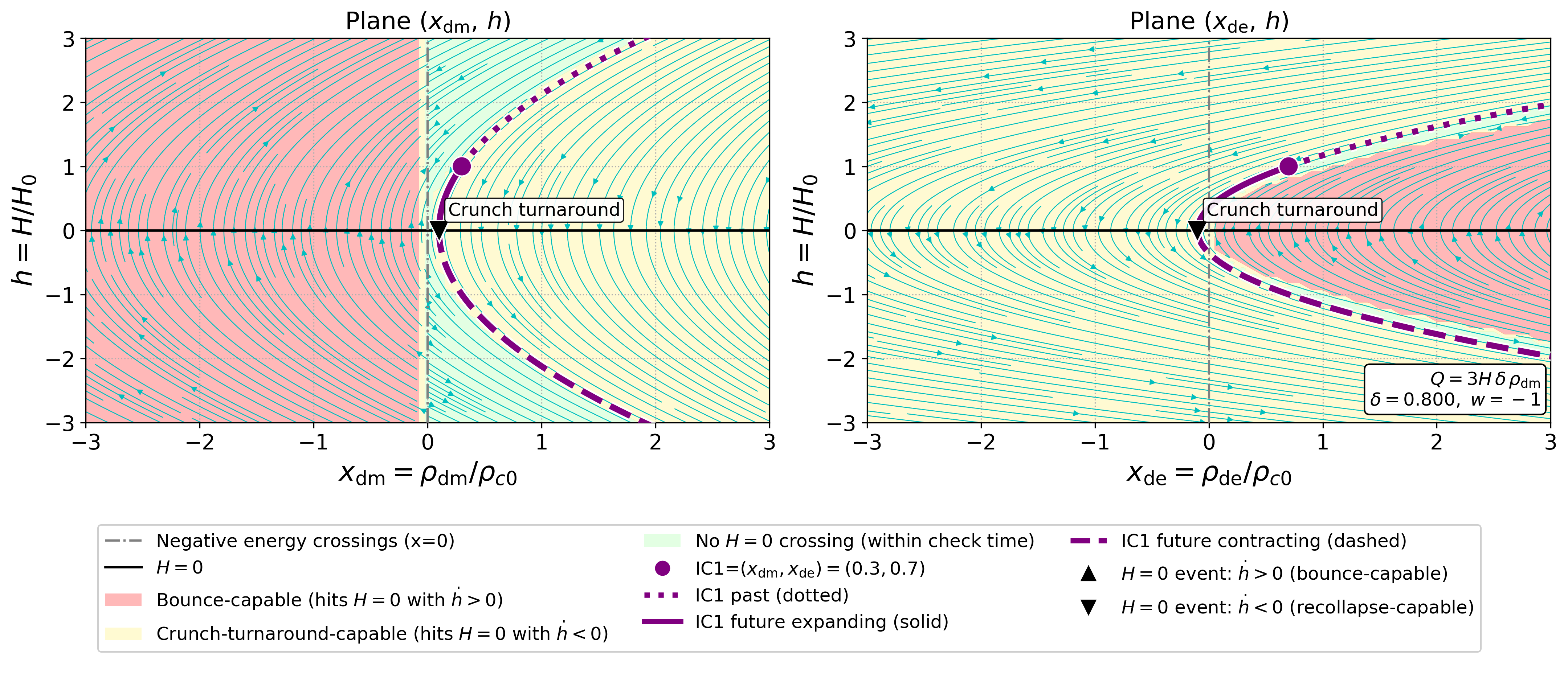}
    \caption{Phase portraits of the dynamical system \eqref{DSA.Qde_w} for interaction kernel $Q= 3 H \delta \rho_{\text{dm}}$, with a large energy transfer from dark energy to dark matter ($\delta\gg0$), which allow for future big crunch-capable turnarounds. The left panel and right panels show dimensionless dark matter $x_{\rm dm}=\rho_{\rm{dm}}/\rho_{(c,0)}$ and dark energy $x_{\rm de}=\rho_{\rm{de}}/\rho_{(c,0)}$ evolution respectively, both relative to the Hubble parameter.  
    The trajectories within the yellow background allow for future big crunch capable turnarounds where $H=0$ and $\dot h<0$, which in the right panel necessitate a crossing into the negative dark energy region ($x_{\rm de}<0$), as required by \eqref{BCdH.5}, and which corresponds to the SiDEDM regime in the left panel of Figure \ref{fig:2D_Q_general_Phase_portrait_boundaries}. Trajectories in the green region do not experience turnarounds and are regions in the iDEDM regime. The bounce-capable turnarounds in the red regions are unphysical, as they correspond to regions where dark matter is negative ($x_{\rm dm}<0$) at present and at all other times, as there exists an invariant submanifold at $x_{\rm dm}=0$ preventing sign-switching for dark matter. The purple dot shows initial conditions in a two fluid model close to those found in Figure \ref{fig:crunch_limit}, \ref{fig:crunch_5} and \ref{fig:2D_Q_general_Phase_portrait_boundaries}. The purple dotted lines show the past expansion, where densities increase approaching the Big Bang. The solid and dashed lines show the future expanding and contracting regions, respectively, indicating dark energy crossing into negative values at the turnaround (here $\rho_{\rm dm}=-\rho_{\rm de}$), before becoming positive again in the contracting phase prior to a crunch. This agrees with Figure \ref{fig:crunch_5} and Table \ref{tab:crunch_phases_ide}.}    \label{fig:DSA_H_Qdm}
\end{figure}

To show that the initial conditions used in Section \ref{sec:crunch_steps} to enable a turnaround is not unique, but one of many such trajectories in the SiDEDM regime, we consider phase portraits described by the dynamical system \eqref{DSA2.7}, derived in Appendix \ref{sec:DSA_new}. This two fluid dynamical system consists of dimensionless variables that are defined at $H=0$:
\begin{gather} \label{DSA2.1_a}
\begin{split}
x_{\rm dm}=\frac{\rho_{\rm{dm}}}{\rho_{(c,0)}} ; \quad x_{\rm de}=\frac{\rho_{\rm{de}}}{\rho_{(c,0)}} ; \quad h=\frac{H}{H_0}.
\end{split}
\end{gather}
Since we assume flatness, we also have the constraint $h^2=x_{\rm dm}^2+x_{\rm de}^2$. For the interaction kernel $Q= 3 H \delta \rho_{\text{dm}}$, we set $\delta_{\rm dm}=\delta$ and $\delta_{\rm de}=0$ in \eqref{DSA2.7}, and in the special case $w=-1$, this system reduces to:

\begin{gather} \label{DSA.Qdm_w}
\begin{split}
 \frac{d x_{\text{dm}}}{d \tau}   = 3h (\delta-1) x_{\text{dm}} ; \quad    \frac{d x_{\text{de}}}{d \tau}   = -3h \delta x_{\text{dm}} ; \quad    \frac{dh}{d \tau}  & =-\frac{3}{2}x_{\rm dm}  .  \\
\end{split}
\end{gather}
Phase portraits in the $(x_{\rm dm},h)$ and $(x_{\rm de},h)$ planes are plotted in Figure \ref{fig:DSA_H_Qdm}. Here we see many trajectories (in the yellow background) that allow for future big crunch capable turnarounds (where $H=0$, $\dot H<0$ and $\rho_{\rm dm}=-\rho_{\rm de}$), with all cases requiring initial conditions that lead to the negative dark energy region ($x_{\rm de}<0$) in the future, as given by condition \eqref{eq:Q_dm_crunch_con} that corresponds to the SiDEDM regime in Figure \ref{fig:2D_Q_general_Phase_portrait_boundaries}. Following the purple trajectory in Figure \ref{fig:DSA_H_Qdm}, densities increase in the past preceding a Big Bang, and dark energy becomes negative in the future before the turnaround at the bounce. This trajectory is symmetric around $H=0$ preceding the contraction phase, where dark energy becomes positive again, and all densities increase approaching a big crunch. This behaviour of the dynamical system \eqref{DSA.Qdm_w} is consistent with the descriptions in Figure \ref{fig:crunch_5} and Table \ref{tab:crunch_phases_ide} from analytical solutions \eqref{eq:rho_dm_Q_dm} and \eqref{eq:rho_de_Q_dm}.

\section{Non-Singular bounce case study: $(Q= 3 H \delta \rho_{\text{de}})$}  \label{sec:bounce}
\subsection{Mathematical background}
\begin{figure}
    \centering
    \includegraphics[width=0.9\linewidth]{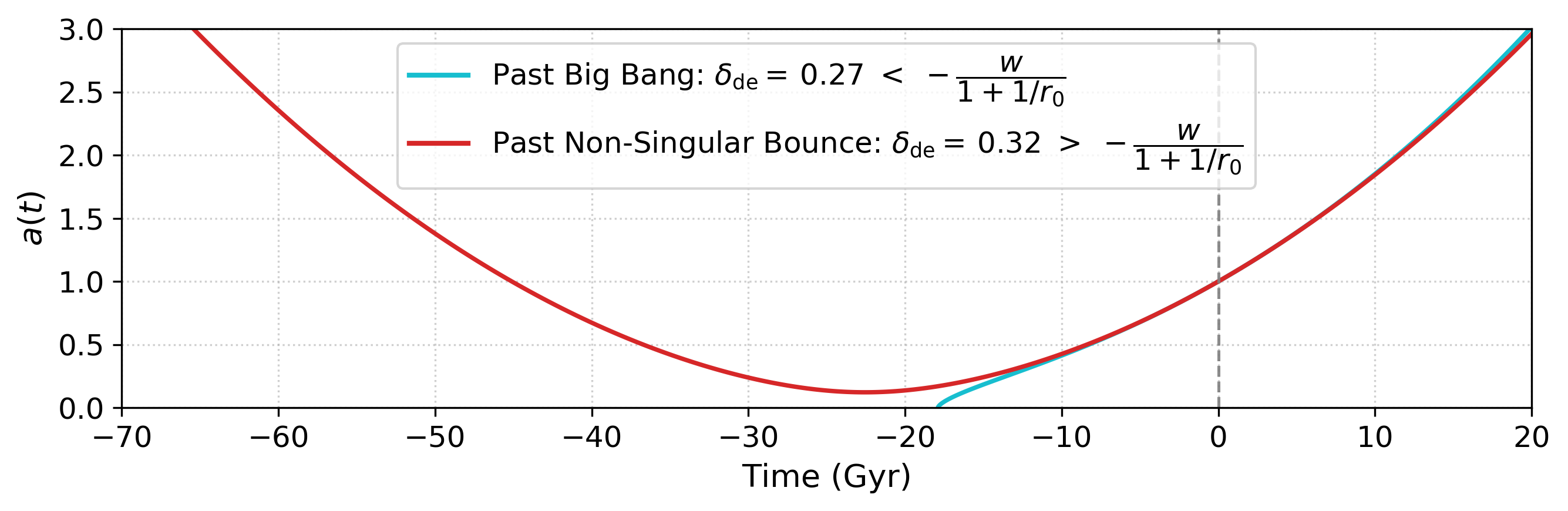}
    \caption{Evolution of scale factor $a$ vs time for interacting dark energy model $Q= 3 H \delta \rho_{\text{de}}$. For both cases we have set $\Omega_{\rm{(bm,0)}}=0.05$, $\Omega_{\rm{(dm,0)}}=0.26$, $\Omega_{\rm{(de,0)}}=0.69$, $w=-1$. From these parameters, the upper positive energy limit is $\delta = -\frac{w }{1+\frac{1}{r_0}}=0.274$. If the interaction strength is below this limit ($\delta =0.270< -\frac{w }{1+\frac{1}{r_0}}$) the universe will experience a past Big Bang, while if above the limit ($\delta =0.320> -\frac{w }{1+\frac{1}{r_0}}$), negative dark matter emerges and the universe will experience a non-singular bounce in the past. 
    A a wide range of initial condition that allow negative dark matter can cause a bounce, as seen in the phase portraits in Figure \ref{fig:DSA_H_Qde}.  The contribution of baryons (included here) causes the need for a slightly larger $\delta$ to observe a bounce.  The negative energy densities emerge from the mechanisms illustrated in Figure \ref{fig:neg_dm_visual} and \ref{fig:2D_Q_general_Phase_portrait_boundaries}.}
    \label{fig:bounce_limit}
\end{figure}

We have observed that the turnaround at a big bounce will also occur when $\rho_{\rm{tot}}=0$. In the models considered here, the past will be dominated by not only dark matter and dark energy, but also baryonic matter and radiation, leading to condition \eqref{BB.1}.
\begin{gather} \label{BB.1}
\begin{split}
\rho_{\rm{tot}}= \rho_{\rm{r}}+\rho_{\rm{bm}}+\rho_{\rm{dm}}+\rho_{\rm{de}}&= 0 \text{ at turnaround } (a_{\rm{min}}). \\
\rho_{\rm{r}}+\rho_{\rm{bm}}+\rho_{\rm{de}}&= -\rho_{\rm{dm}} \text{ at turnaround } (a_{\rm{min}}). \\
\end{split}
\end{gather} 
This condition can only hold in a flat universe if one of the components, for example $\rho_{\rm{dm}}$ is negative in the past.
IDE models experience a non-singular big bounce when there is a large energy transfer from dark energy to dark matter, corresponding to energy flow from dark matter to dark energy as we move further back into the past, until DM becomes negative and \eqref{BB.1} holds. For the special case where $\delta_{\rm dm}=0$, we have the popular interaction kernel $Q= 3 H \delta \rho_{\text{de}}$, for which the dark matter and dark energy evolves with the following equations  \cite{vanderWesthuizen:2025I}:
\begin{gather}
\begin{split} \label{eq:rho_dm_Q_de}
\rho_{\text{dm}} =  \left(\rho_{\text{(dm,0)}}  + \rho_{\text{(de,0)}} \left(\frac{\delta}{\delta+w} \right) \left[1-  a^{-3(\delta + w)} \right] \right) a^{-3},
\end{split}   
\end{gather}
\begin{gather}
\begin{split} \label{eq:rho_de_Q_de}
\rho_{\text{de}} = \rho_{\text{(de,0)}} a^{-3(\delta + w + 1)}.
\end{split}   
\end{gather}
We can see that $\rho_{\rm de}$ will always be positive, while $\rho_{\rm dm}$ may become negative, so from the conditions in \eqref{BCdH.5} a turnaround will be associated with a bounce for this model, as it cannot experience a crunch. We will now see that the strongly interacting case will correspond physically to a bounce and eventual re-expansion on the past, as stated in \eqref{eq:Q_dm_bounce_con}.  The reason for this is that this upper limit is associated with negative dark matter in the past, which has a repulsive anti-gravity effect that can cause an eventual re-expansion of the universe. This repulsive effect of negative dark matter has been noted in \cite{Bondi:1957zz, Najera:2021tcx, Manfredi:2026qoz} can be seen from the Raychaudhuri equation \eqref{F2} in natural units, while keeping in mind $p_{\rm{dm}}=0$ and $w_{\rm{dm}}=0$, as shown in \eqref{eq:dm_neg_repulsive}.
\begin{equation}
\begin{split}
\text{Negative DM is repulsive}: \underbrace{\bigg( \frac{\ddot{a}}{a}\bigg)}_{>0}  &=\underbrace{ - \frac{4\pi G}{3}  \left[ \underbrace{\rho_{\rm{dm}}}_{<0} \underbrace{(1+  3w_{\rm{dm}})}_{>0} \right]}_{>0}.
\end{split}
\label{eq:dm_neg_repulsive}
\end{equation}
The condition for a non-singular bounce to occur in the past is \eqref{eq:Q_dm_bounce_con}, derived in Appendix \ref{sec:mimmax} from the requirement of a positive and real solution for $a_{\rm min/max}$ in \eqref{eq:rhodm=de_dm}, and illustrated in Figure \ref{fig:bounce_limit}. The condition \eqref{eq:Q_dm_bounce_con} corresponds to the violation of the upper positive energy bound for energy transfer from dark energy to dark matter found in \cite{vanderWesthuizen:2023hcl, vanderWesthuizen:2025I}.
\begin{equation}
\begin{split}
\text{Flat Universe Non-Singular Bounce Condition : }\quad \delta > -\frac{w }{\left(1+\frac{1}{r_0}\right)} \quad  \text{(SiDEDM regime)}. 
\end{split}
\label{eq:Q_dm_bounce_con}
\end{equation}
It should be mentioned that this condition is only strictly valid in a two-fluid universe model containing only dark matter and dark energy. The addition of baryons and radiation to the model will cause an extra attractive gravitational effect, for which a larger $\delta$ is needed to create the larger repulsive effect needed to overcome gravity as required for a bounce to occur.  

As we move into the past from the present, a turnaround from expansion to contraction will occur when the universe reaches its minimum size $a_{\rm{max}}$ when $\rho_{\rm{r}}+\rho_{\rm{bm}}+\rho_{\rm{de}}= -\rho_{\rm{dm}}$. An analytical expression for $a_{\rm{min}}$ can only be obtained in the simplifying cases when radiation or baryonic matter are assumed to be negligible during the bounce. Thus, substituting \eqref{eq:rho_dm_Q_de} and \eqref{eq:rho_de_Q_de} into \eqref{BB.1}, setting $\rho_{\rm{r}}+\rho_{\rm{bm}} \approx 0$ and solving for $a$, yields:
\begin{equation}
\begin{split}
a_{\rm{min}}  =& \left[-\frac{   \Omega_{\text{(dm,0)}} }{ \Omega_{\text{(de,0)}}  } \left( \frac{  \delta+w }{w} \right) -\frac{\delta}{w}\right]^{-\frac{1}{{3(\delta +w) }}} .
\end{split}
\label{eq:rhodm=de_dm}
\end{equation}
A more realistic case can be obtained if we take only radiation to be negligible $\rho_{\rm{r}}=0$ and solve for $\rho_{\rm{bm}}+\rho_{\rm{de}}= -\rho_{\rm{dm}}$:
\begin{equation}
\begin{split}
a_{\rm{min}}  =&\left[- \frac{\Omega_{\text{(bm,0)}}+  \Omega_{\text{(dm,0)}}}{\Omega_{\text{(de,0)}}} \left(\frac{\delta+w}{w} \right) - \frac{\delta}{w} \right]^{\frac{1}{-3(\delta + w )}}. \\
\end{split}
\label{eq:a_min_1}
\end{equation}
Both \eqref{eq:rhodm=de_dm} and \eqref{eq:a_min_1} are derived in Appendix \ref{sec:mimmax}. We can see that this approximation holds when we substitute in the parameters $\Omega_{\rm{(dm,0)}}=0.26$, $\Omega_{\rm{(de,0)}}=0.69$, $\Omega_{\rm{(bm,0)}}=0.05$ , $w=-1$, and $\delta=+0.34$ into \eqref{eq:a_min_1}, which gives $a_{\rm{min}}=0.205$. This matches the values for $a_{\rm{min}}$ seen in the top panel of Figure \ref{fig:bounce_5}, which is created with the same parameters. We can see that our model closely resembles a closed de Sitter universe. To determine the likelihood of a past non-singular bounce, we need to obtain observational constraints on the coupling parameter $\delta$ that obeys the
condition in \eqref{eq:Q_dm_bounce_con}. For this exact model, recent late-time background constraints were obtained in \cite{Figueruelo2026IDEconstraints}, which resulted in posteriors on $\delta$ that allow for \eqref{eq:Q_dm_bounce_con} within both the 65\% and 95\% confidence intervals, thus allowing the behaviour, but making it unlikely (see Figure 6 in \cite{Figueruelo2026IDEconstraints}, which may be compared with Figure \ref{fig:Q_parameter} in this work).

\subsection{Step-by-step non-singular bounce mechanism}  \label{sec:bounce_steps}

\begin{figure}
    \centering
    \includegraphics[width=0.9\linewidth]{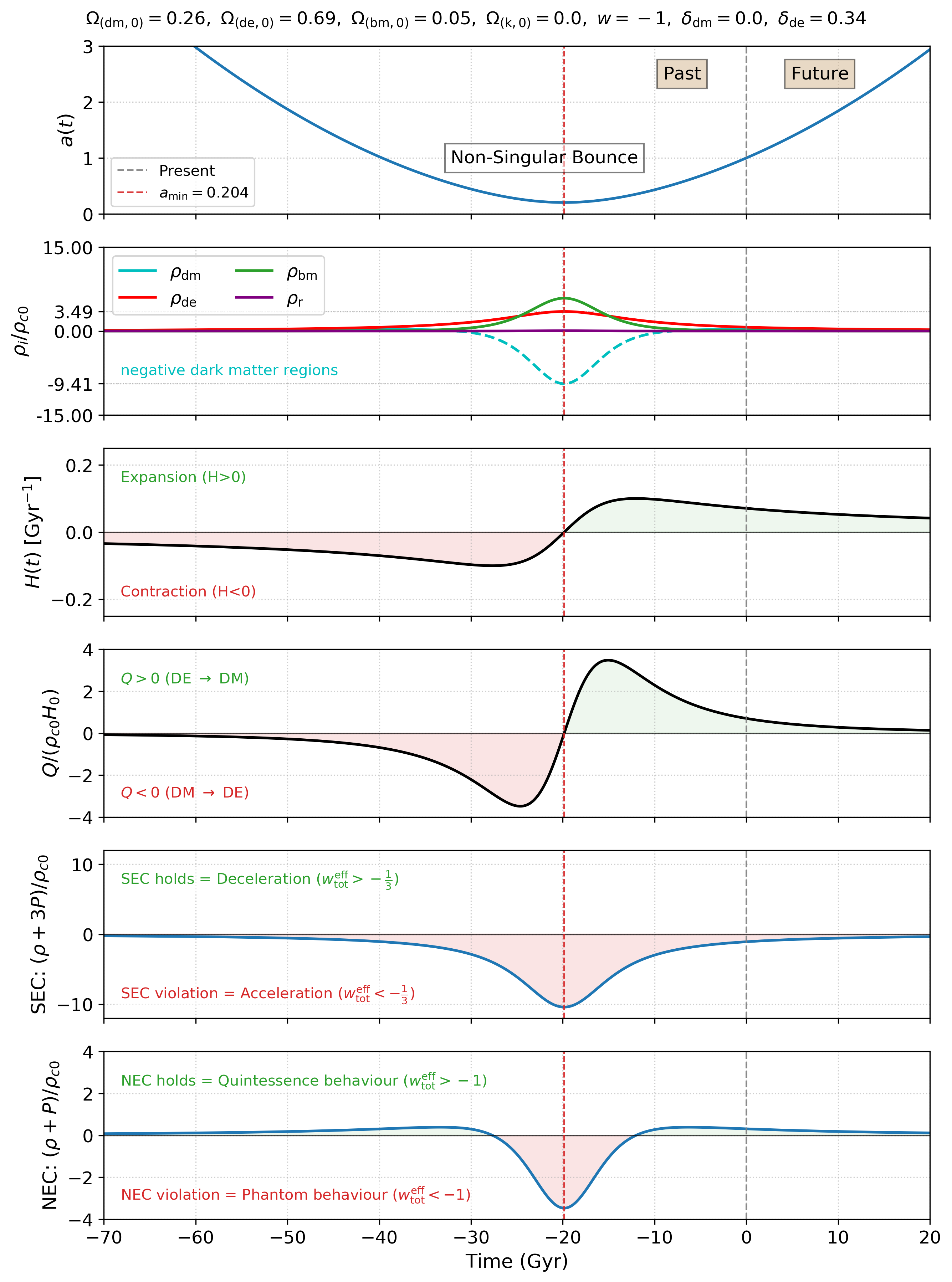}{}
    \caption{ Evolution of the scale factor 
    $a(t)$, dimensionless energy densities $\rho_i/\rho_{c0}$, Hubble parameter $H(t)$, dimensionless interaction term $Q/(\rho_{c0}H_0)$, and the strong and null energy conditions (SEC and NEC) for interacting dark energy model $Q= 3 H \delta \rho_{\text{de}}$. Energy is transferred from DE to DM at present, corresponding to energy flow from DM to DE as we move backwards in time.  As illustrated in Figure \ref{fig:neg_dm_visual}, DM becomes negative in the past, since $Q>0$ when $\rho_{\rm{dm}}=0$. At some point $\rho_{\rm{r}}+\rho_{\rm{bm}}+\rho_{\rm{de}}= -\rho_{\rm{dm}}$, which causing expansion to stop as $H=0$, which in return causes $Q=0$. This point is a non-singular ($a\neq0$) bounce, which is associated with the violation of the strong and null energy conditions (bottom two panels). Moving further into the past we see a period of contraction before the bounce, where DM is positive again. See Figure \ref{fig:DSA_H_Qde} for phase portraits showing similar behaviour.}
    \label{fig:bounce_5}
\end{figure}

We will now consider a similar analysis as done in section \ref{sec:crunch_steps}, but for a non-singular bounce. We will again start with initial conditions close to that of $\Lambda$CDM, with a present era of accelerated expansion. In this case we have added a strong energy transfer ($\delta_{\rm de}=0.34$) from dark energy to dark matter, as required by \eqref{eq:Q_dm_bounce_con} for a past non-singular bounce in a flat universe. 
In Figure \ref{fig:bounce_5}, we plot the same parameters as in Figure \ref{fig:crunch_5} against time. The plots are made using the two Friedmann equations \eqref{F1} and \eqref{F2}, the evolution of the energy densities \eqref{eq:rho_dm_Q_de} and \eqref{eq:rho_de_Q_de}, and the energy conditions \eqref{1.3.1} and \eqref{BCdH.3}. Our analysis follows Figure \ref{fig:bounce_5} from right to left, starting from the present and moving further into the past, mentioning important transitions along the way.

\begin{enumerate}
    \item \underline{Present accelerating expansion ($a=1 ;\;\rho_{\rm dm}>0;\; H>0 ; \; Q>0 \;(\text{DE}\rightarrow\text{DM}); \; \text{SEC violated; NEC holds}$)}: \\ \\
    The model starts with initial conditions similar to that of the flat $\Lambda$CDM model, with all energies positive, the NEC holds, while the SEC is violated, giving the present era of accelerated expansion. The only modification is a strong coupling $Q\gg0$ that causes energy transfer from DE to DM, with $\delta$ exceeding the limit derived in \eqref{eq:Q_dm_bounce_con}.
    
    \item \underline{Negative DM expansion ($a_{\rm{min}}<a<1 ;\;\rho_{\rm dm}<0;\; H>0 ; \; Q>0 \;(\text{DE}\rightarrow\text{DM}); \; \text{SEC \& NEC violated} $)}: \\ \\
    The large energy transfer from DE to DM corresponds to energy transfer from DM to DE as we move from the present into the past. This energy transfer depletes DM in the past, which eventually becomes negative ($\rho_{\rm dm}<0$) and repulsive as shown in \eqref{eq:dm_neg_repulsive}, further accelerating the expansion rate until $w_{\rm tot}^{\rm eff}<-1$ and the NEC is violated.

    \item \underline{Bounce at minimum scale factor ($a=a_{\rm{min}} ;\;\rho_{\rm dm}<0;\; H=0 ; \; Q=0 \;(\text{no transfer}); \; \text{SEC \& NEC violated} $)}: \\ \\
    The dark matter component becomes negative enough so that $\rho_{\rm tot}=0$ as given in condition \eqref{BB.1}. From the first Friedmann equation \eqref{F1}, this leads to $H=0$ and the expansion to stop at a minimum non-singular scale factor ($a_{\rm min}\neq 0$) determined by \eqref{eq:a_min_1}. From \eqref{F2}, we still have $\dot{H}>0$, as shown by the SEC violation, which implies acceleration throughout the turnaround, which will therefore be preceded by a period of contraction in the past before the bounce. Similar to the turnaround for the big crunch, we again have a breakdown of the total effective equation of state $w^{\rm eff}_{\rm tot}=\frac{P_{\rm{tot}}}{\rho_{\rm{tot}}} =-\infty$.

    \item \underline{Negative DM accelerated contraction ($a>a_{\rm{min}} ;\;\rho_{\rm dm}<0;\; H<0 ; \; Q<0 \;(\text{DM}\rightarrow\text{DE}); \; \text{SEC \& NEC violated} $)}: \\ \\
    Before the bounce, as the universe was contracting, the direction of energy transfer switched ($H<0$ and $Q \propto H$) and energy flows from DM to DE (corresponding to energy flow from DE to DM moving further into the past). This energy transfer continues and eventually the DM component becomes positive in the more distant past. This also causes an end to the NEC violation. This final phase continues indefinitely into the past.

    \item \underline{Positive DM accelerated contraction ($a\gg1 ;\;\rho_{\rm dm}>0;\; H<0 ; \; Q<0 \;(\text{DM}\rightarrow\text{DE}); \; \text{SEC violated; NEC holds}$)}: \\ \\
    In this final phase, all components have positive energy and the universe will expand eternally into the past, similar to a closed de Sitter model. In this case, the expansion dilutes all components such that $\rho_{\rm i}\rightarrow0$ as $t\rightarrow -\infty$.

\end{enumerate}

In the steps above, the required bounce conditions from Table \ref{tab:bounce_flat_con} are obeyed. The main features of the IDE bounce shown in Figure \ref{fig:bounce_5} is summarised in Table \ref{tab:bounce_phases_ide}.

\begin{table}[H]
\centering
\renewcommand{\arraystretch}{1.25}
\begin{tabular}{|c|c|c|c|c|c|c|}
\hline
\textbf{Phase} 
& $\boldsymbol{a}$ 
& $\boldsymbol{H}$ 
& $\boldsymbol{\rho_{\rm dm}}$ 
& $\boldsymbol{Q}$ 
& \textbf{SEC} 
& \textbf{NEC} \\
\hline
\hline
Present accelerating expansion 
& $1$ 
& $+$ 
& $+$ 
& $>0$ (DE$\rightarrow$DM) 
& Violated ($w^{\rm eff}_{\rm tot}<-\frac{1}{3}$) 
& Holds \\ 
\hline
Negative DM accelerating expansion 
& $a_{\rm min}<a<1$ 
& $+$ 
& $-$ 
& $>0$ (DE$\rightarrow$DM) 
& Violated ($w^{\rm eff}_{\rm tot}<-\frac{1}{3}$) 
& Violated ($w^{\rm eff}_{\rm tot}<-1$) \\ 
\hline
Bounce (minimum scale factor) 
& $a_{\rm min}$ 
& $0$ 
& $-$ 
& $0$ (no transfer) 
& Violated ($\dot{H}>0$) 
& Violated ($w^{\rm eff}_{\rm tot}<-1$) \\ 
\hline
Negative DM accelerated contraction 
& $a>a_{\rm min}$ 
& $-$ 
& $-$ 
& $<0$ (DM$\rightarrow$DE) 
& Violated ($w^{\rm eff}_{\rm tot}<-\frac{1}{3}$)
& Violated ($w^{\rm eff}_{\rm tot}<-1$) \\ 
\hline
Positive DM accelerated contraction 
& $a\gg1$ 
& $-$ 
& $+$ 
& $<0$ (DM$\rightarrow$DE) 
& Violated ($w^{\rm eff}_{\rm tot}<-\frac{1}{3}$) 
& Holds \\ 
\hline
\end{tabular}
\caption{Sequence of cosmological phases, from the present (top row) and moving further into past with each descending row, describing a non-singular bounce in the interacting dark energy model. The bounce occurs at a finite minimum scale factor $a_{\rm min}\neq0$ where $H=0$ and $\dot H>0$. The sign of the interaction term $Q$ indicates the direction of energy transfer between dark energy (DE) and dark matter (DM).}
\label{tab:bounce_phases_ide}
\end{table}

\subsection{Bounce mechanism from phase portraits}
\begin{figure}
    \centering
    \includegraphics[width=0.95 \linewidth]{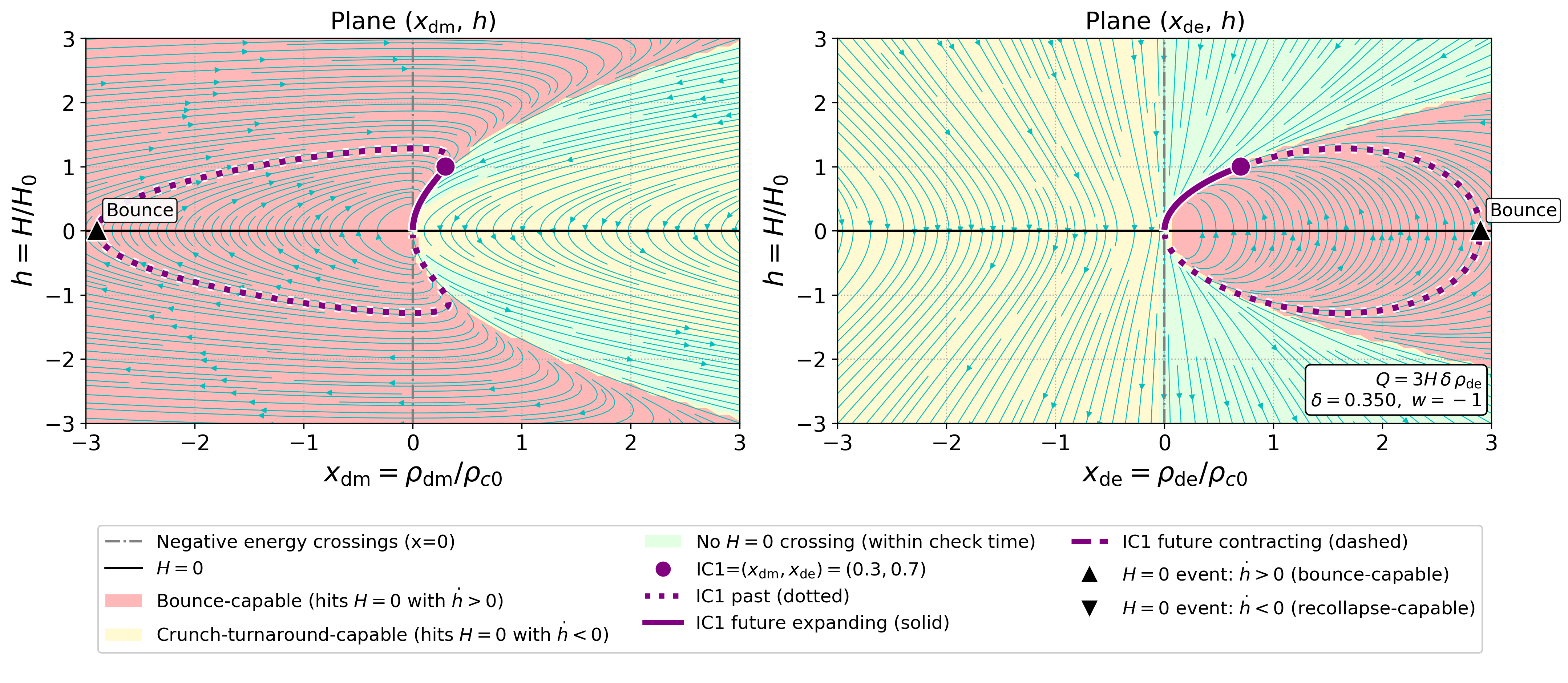}
    \caption{Phase portraits of the dynamical system \eqref{DSA.Qde_w} for interaction kernel $Q= 3 H \delta \rho_{\text{de}}$, with a large energy transfer from dark energy to dark matter ($\delta\gg0$), which allow for past big bounce turnarounds. The left panel and right panels show dimensionless dark matter $x_{\rm dm}=\rho_{\rm{dm}}/\rho_{(c,0)}$ and dark energy $x_{\rm de}=\rho_{\rm{de}}/\rho_{(c,0)}$ evolution respectively, both relative to the Hubble parameter. The trajectories within the red background allow big bounce turnarounds where $H=0$ and $\dot h>0$, which in the left panel necessitate a crossing into the negative dark matter region ($x_{\rm de}<0$) in the past, as required by \eqref{BCdH.5}, and which corresponds to the SiDEDM regime in the right panel of Figure \ref{fig:2D_Q_general_Phase_portrait_boundaries}. Trajectories in the green region do not experience turnarounds and are regions in the iDEDM regime. The crunch-capable turnarounds in the yellow regions are unphysical, as they corresponds to region where dark energy is negative ($x_{\rm de}<0$) at present and at all other times, as there exists an invariant submanifold at $x_{\rm de}=0$ preventing sign-switching for dark energy. The purple dot shows initial conditions in a two fluid model similar to those found in Figure \ref{fig:bounce_limit}, \ref{fig:bounce_5} and \ref{fig:2D_Q_general_Phase_portrait_boundaries}, with the dotted lines showing the past expansion where dark matter crosses into negative values prior to the bounce (here $-\rho_{\rm dm}=\rho_{\rm de}$), before becoming positive again in contracting phase. The solid purple lines shows future expansion where $h\rightarrow0$ as $\rho_{\rm dm}\rightarrow0$ and $\rho_{\rm de}\rightarrow0$, which is also observed in the end of the dotted line in the contracting phase. The behaviour described here is in agreement with Figure \ref{fig:bounce_5}.}    \label{fig:DSA_H_Qde}
\end{figure}

We will now show that the observed bouncing behaviour may also be obtained from a dynamical system analysis. For the interaction kernel $Q= 3 H \delta \rho_{\text{de}}$, we set $\delta_{\rm dm}=0$ and $\delta_{\rm de}=\delta$ in \eqref{DSA2.7}, and in the special case $w=-1$, this system reduces to:

\begin{gather} \label{DSA.Qde_w}
\begin{split}
 \frac{d x_{\text{dm}}}{d \tau}   = 3h\left[- x_{\text{dm}} + \delta_{\text{de}}  x_{\text{de}}\right]; \quad   \frac{d x_{\text{de}}}{d \tau}   = -3h \delta_{\text{de}}  x_{\text{de}}; \quad   \frac{dh}{d \tau}   =-\frac{3}{2}x_{\rm dm} .  
\end{split}
\end{gather}
Phase portraits in the $(x_{\rm dm},h)$ and $(x_{\rm de},h)$ planes are plotted in Figure \ref{fig:DSA_H_Qde}. Here we see many trajectories (in the red background) that allow for past bounce turnarounds (where $H=0$, $\dot H>0$ and $-\rho_{\rm dm}=\rho_{\rm de}$), with all cases requiring initial conditions that lead to the negative dark matter region ($x_{\rm dm}<0$) in the past, as given by condition \eqref{eq:Q_dm_bounce_con} that corresponds to the SiDEDM regime also shown in Figure \ref{fig:2D_Q_general_Phase_portrait_boundaries}. Following the purple trajectory in Figure \ref{fig:DSA_H_Qde}, we see that the trajectory predicts an eternal future expansion where all fluids dilute, while negative dark matter densities emerge in the past prior to the bounce. The trajectory is also symmetric around the bounce at $H=0$, thus allowing dark matter to become positive again  and all fluids to dilute as we approach the more distant past in the contraction phase. These features suggest the behaviour of our dynamical system \eqref{DSA2.7} is in agreement with the descriptions obtained using the analytical solutions \eqref{eq:rho_dm_Q_de} and \eqref{eq:rho_de_Q_de}, which resulted in Figure \ref{fig:bounce_5} and Table \ref{tab:bounce_phases_ide}.

\section{Cosmological turnarounds in other phenomenological IDE models} \label{sec:otherQ}

In this study, we have chosen to focus on the two simplest interaction kernels $Q=3 H \delta \rho_{\text{dm}}$ and $Q= 3 H \delta \rho_{\text{de}}$, as they most clearly elucidate the physical processes that may lead to a big crunch or non-singular bounce. Non-linear interaction kernels like  $Q_2=3\delta H  
\left(\frac{\rho_{\text{dm}}^2}{\rho_{\text{dm}}+\rho_{\text{de}}} \right)$ and $Q_2=3\delta H  
\left(\frac{\rho_{\text{de}}^2}{\rho_{\text{dm}}+\rho_{\text{de}}} \right)$ have similar background dynamics, but a big crunch or non-singular bounce cannot be obtained since the upper limit on $\delta$ in the iDEDM regime of both models is plagued by undefined energy densities (divisions by zero, see Table 1 and the bottom two panels of Figure 1 in \cite{vanderWesthuizen:2025III}). This so called strong coupling SiDEDM regime can therefore not be reached. Conversely, the non-linear interaction $Q_1=3\delta H  
\left(\frac{\rho_{\text{dm}}\rho_{\text{de}}}{\rho_{\text{dm}}+\rho_{\text{de}}} \right)$ always has positive energies, and therefore a turnaround in a flat universe cannot occur.

The linear interactions $Q= 3 H (\delta_{\text{dm}} \rho_{\text{dm}} + \delta_{\text{de}}  \rho_{\text{de}})$ and $Q=3H\delta( \rho_{\text{dm}}+\rho_{\text{de}})$ allow the same crunching and bouncing behaviour, but in the first case this introduces another free parameter, while in both cases the overall parameter space is more complicated, as imaginary energy densities appear near the upper SiDEDM regime (see Table 1 in \cite{vanderWesthuizen:2025III}). These models can show the same behaviour, but in a more complicated manner and do not add much new to the discussion. For this reason, we have only included Figure \ref{fig:crunch_bounce_limit}, which shows how the turnaround condition \eqref{DSA.Q.dm+de.PEC} derived in Appendix \ref{sec:mimmax} for $Q=3H\delta( \rho_{\text{dm}}+\rho_{\text{de}})$, may induce a non-singular bounce in the past (if $\Omega_{\rm{(dm,0)}}<\Omega_{\rm{(de,0)}}$ or equivalently $r_0<1$) or a big crunch (if $\Omega_{\rm{(dm,0)}}>\Omega_{\rm{(de,0)}}$ or equivalently $r_0>1$). 
\begin{gather} \label{DSA.Q.dm+de.PEC}
\text{Flat Universe Bounce/Crunch Condition : } \;1. \;\delta > -\frac{w r_0}{(1+r_0)^2}  \quad ;  \quad 2. \; \delta \le -\frac{w}{4}  \quad \text{(SiDEDM regime)}.
\end{gather}
In both cases, the minimum or maximum scale factor is given by \eqref{eq:a_min_max_dm+de}, with the predicted values shown by the red dashed lines in Figure \ref{fig:crunch_bounce_limit}.
\begin{equation}
\begin{split}
a_{\rm{min/max}}
= \left[ 
\frac{
\Omega_{\text{(de,0)}} ( w + \Delta) + \Omega_{\text{(dm,0)}} ( - w + \Delta)
}{
\Omega_{\text{(de,0)}} ( w - \Delta) + \Omega_{\text{(dm,0)}} ( - w - \Delta)
}
\right]^{\frac{1}{3 \Delta}}.
\end{split}
\label{eq:a_min_max_dm+de}
\end{equation}
The minimum and maximum scale factor and turnaround conditions for some other linear interaction kernels can be found in the appendix \ref{sec:mimmax}.

\begin{figure}
    \centering
    \includegraphics[width=0.9\linewidth]{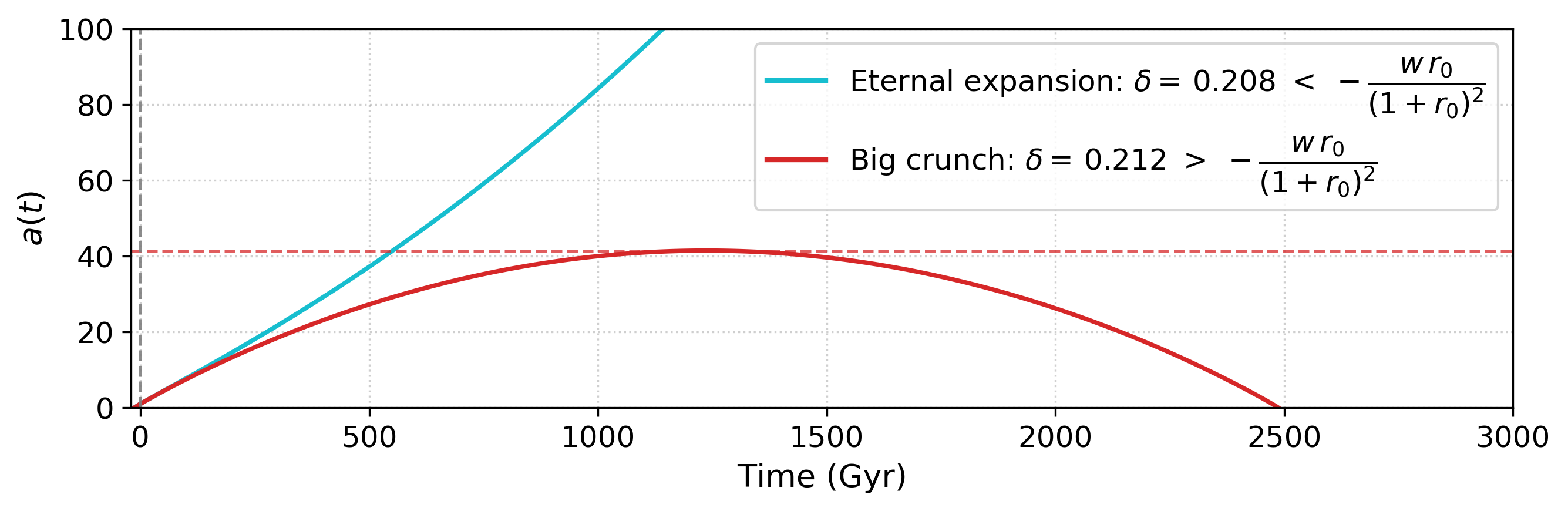}
    \includegraphics[width=0.9\linewidth]{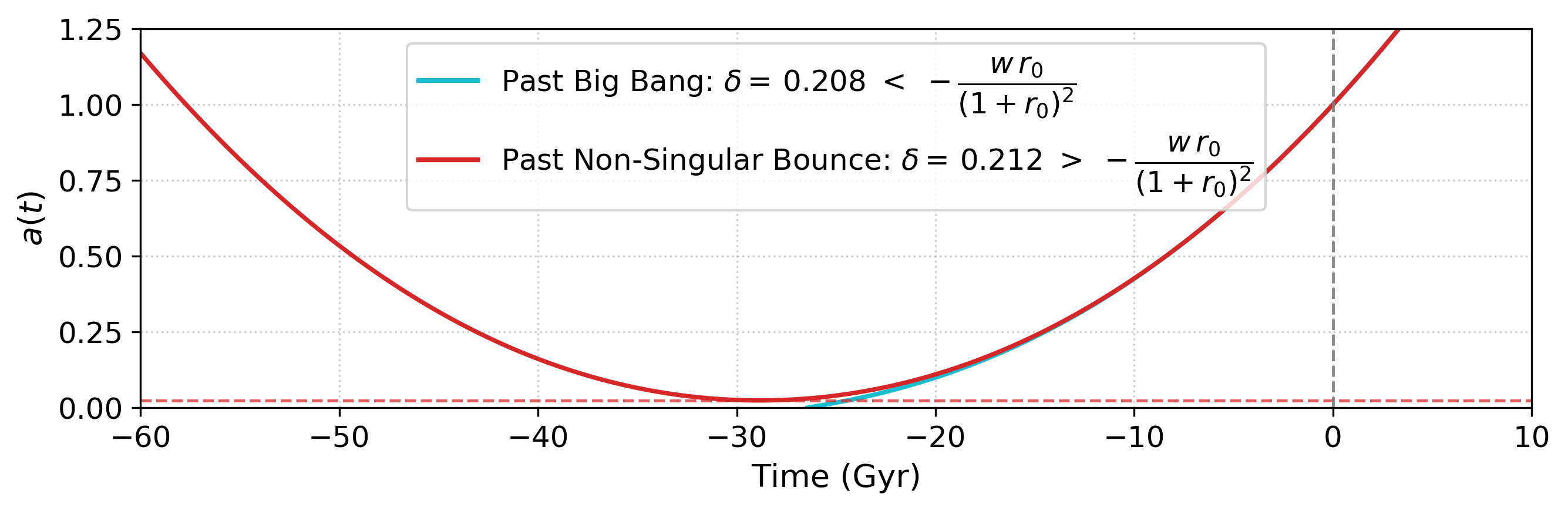}
    \caption{Evolution of scale factor $a$ vs time for interacting dark energy model $Q=3H\delta( \rho_{\text{dm}}+\rho_{\text{de}})$. For both cases we have set $\Omega_{\rm{(r,0)}}=\Omega_{\rm{(bm,0)}}=0.0$ and $w=-1$, while for the top panel we have $\Omega_{\rm{(dm,0)}}=0.7$, $\Omega_{\rm{(de,0)}}=0.3$, and the bottom panel $\Omega_{\rm{(dm,0)}}=0.3$, $\Omega_{\rm{(de,0)}}=0.7$ . From these parameters, for both cases the upper positive energy limit is $\delta = -\frac{w r_0}{(1+r_0)^2}=0.210$. If the interaction strength is below this limit ($\delta =0.208< -\frac{w r_0}{(1+r_0)^2}$) the universe will experience a past Big Bang and future eternal expansion, while if above the limit ($\delta =0.212> -\frac{w r_0}{(1+r_0)^2}$) the universe will experience either a non-singular bounce in the past (if $\Omega_{\rm{(dm,0)}}<\Omega_{\rm{(de,0)}}$ or equivalently $r_0<1$) or a big crunch (if $\Omega_{\rm{(dm,0)}}>\Omega_{\rm{(de,0)}}$ or equivalently $r_0>1$). The dashed red lines show the predicted values for $a_{\rm{min/max}}$ using \eqref{eq:a_min_max_dm+de}, which match with the plots for $a$. Baryons and radiation have been neglected here to show that the limit is for a two fluid model. Including additional fluids will cause a slightly larger $\delta$ to observe a bounce or crunch. A phase portrait showing how negative energy emerges for the interaction kernel is given in Figure \ref{fig:2D_Q_dm+de_Phase_portrait_boundaries}, while crunching and bouncing can also be seen in Figure \ref{fig:DSA_H_Qdm+de}.}
    \label{fig:crunch_bounce_limit}
\end{figure} 

\begin{figure}
    \centering
    \includegraphics[width=0.95 \linewidth]{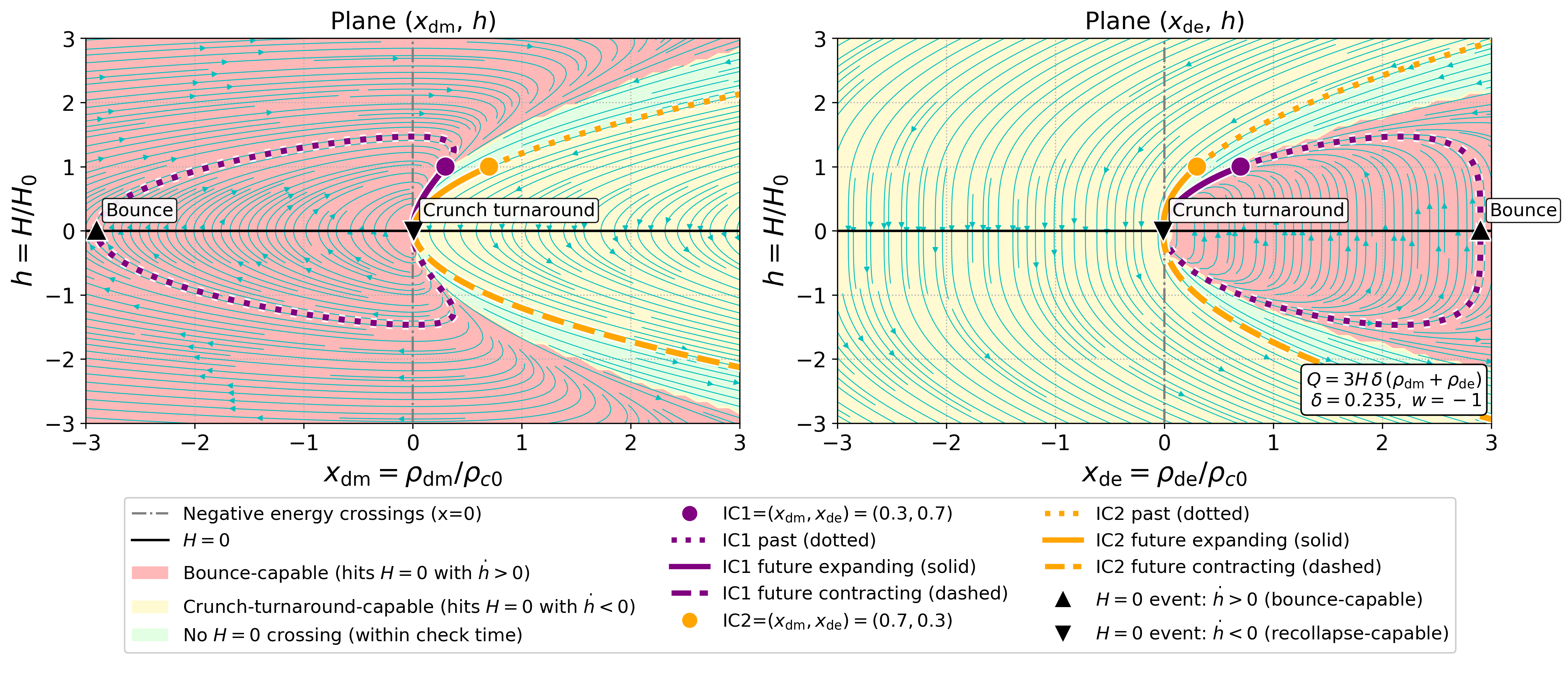}
    \caption{Phase portraits of the dynamical system \eqref{DSA.Qdm+de_w} for interaction kernel $Q= 3 H \delta(\rho_{\text{dm}} +\rho_{\text{de}})$, with a large energy transfer from dark energy to dark matter ($\delta\gg0$), which allow for both future crunch-capable and past big bounce turnarounds, depending on the initial conditions. The left panel and right panels show dimensionless dark matter $x_{\rm dm}=\rho_{\rm{dm}}/\rho_{(c,0)}$ and dark energy $x_{\rm de}=\rho_{\rm{de}}/\rho_{(c,0)}$ evolution respectively, both relative to the Hubble parameter. The trajectories within the red background allow big bounce turnarounds where $H=0$ and $\dot h>0$, which in the left panel necessitate a crossing into the negative dark matter region ($x_{\rm de}<0$) in the past, as required by \eqref{BCdH.5}. The purple dot corresponds to initial conditions where $r_0<1$ in the SiDEDM regime shown in Figure \ref{fig:2D_Q_dm+de_Phase_portrait_boundaries}, while the purple dotted, solid and dashed lines behave as described in Figure \ref{fig:DSA_H_Qde}, allowing a future crunch.
    Similarly, the trajectories within the yellow background allow for future big crunch capable turnarounds where $H=0$ and $\dot h<0$, which in the right panel necessitate a crossing into the negative dark energy region ($x_{\rm de}<0$), as required by \eqref{BCdH.5}. The orange dot corresponds to initial conditions where $r_0>1$ in the SiDEDM regime shown in Figure \ref{fig:2D_Q_dm+de_Phase_portrait_boundaries}, while the solid dotted, solid and dashed lines behave as described in Figure \ref{fig:DSA_H_Qdm}, allowing a bounce. This single interaction allowing both bouncing and crunching behaviour, depending on the value of $r_0$, is in agreement with Figure \ref{fig:crunch_bounce_limit}.}    \label{fig:DSA_H_Qdm+de}
\end{figure}

The presence of both a crunch and bounce for the same interaction, depending on the value of $r_0$, can also be obtained from the dynamical system considerations. For the interaction kernel $Q= 3 H \delta \rho_{\text{de}}$, we set $\delta_{\rm dm}=0$ and $\delta_{\rm de}=\delta$ in \eqref{DSA2.7}, and in the special case $w=-1$, this system reduces to:

\begin{gather} \label{DSA.Qdm+de_w}
\begin{split}
 \frac{d x_{\text{dm}}}{d \tau}   = 3h\left[ (\delta-1) x_{\text{dm}} + \delta  x_{\text{de}}\right]; \quad   \frac{d x_{\text{de}}}{d \tau}   = -3h \delta( x_{\text{dm}} + x_{\text{de}}); \quad   \frac{dh}{d \tau}   =-\frac{3}{2}x_{\rm dm} .  
\end{split}
\end{gather}
Phase portraits of system \eqref{DSA.Qdm+de_w} are plotted in Figure \ref{fig:DSA_H_Qdm+de}. We see both trajectories (in the red background) that allow for past bounce turnarounds requiring negative dark matter ($x_{\rm dm}<0$) in the past, while trajectories (in the yellow background) allow for future big crunch capable turnarounds (where $H=0$, $\dot H<0$ and $\rho_{\rm dm}=-\rho_{\rm de}$) requiring negative dark energy ($x_{\rm de}<0$) in the future. The red and yellow region both correspond to the SiDEDM regime given by condition \eqref{DSA.Q.dm+de.PEC_BG} with $r_0<1$ and $r_0>1$ respectively. Both of these caes are also illustrated in Figure \ref{fig:2D_Q_dm+de_Phase_portrait_boundaries}.

\section{Summary of Results and Conclusions} \label{sec:disc}

In this work, we have studied the consequences of strong non-gravitational interactions within the dark sector, in which there is a large energy transfer from dark energy to dark matter, $Q\gg0$. In this regime, one or both dark-sector components can cross into negative energy densities. As discussed in Section \ref{sec:neg}, these negative energies arise because phenomenological interaction models often allow a non-vanishing energy transfer $Q\neq0$ even when $\rho_{\rm dm}=0$ or $\rho_{\rm de}=0$, permitting the densities to evolve through zero. This behaviour is illustrated intuitively for a volume element in Figures \ref{fig:neg_de_visual} and \ref{fig:neg_dm_visual}, while phase portraits demonstrating the absence of invariant submanifolds at the zero-density boundaries are shown in Figure \ref{fig:2D_Q_general_Phase_portrait_boundaries}. Although these features were noted previously in \cite{vanderWesthuizen:2025I,vanderWesthuizen:2025II,vanderWesthuizen:2025III}, their impact on the background cosmological dynamics had not been explored in detail. Here we fill this gap by explicitly demonstrating how such interactions can give rise to crunching, bouncing, and cyclic cosmologies at the background level.

In Section \ref{sec:Conditions_general}, we derived the background-level conditions within general relativity required for cosmological turnarounds associated with either a crunch or a bounce, with particular emphasis on the violation of the strong and null energy conditions. These conditions are shown to be satisfied for a future crunch in the model $Q=3H\delta\rho_{\rm dm}$ in Section \ref{sec:crunch}, and for a past non-singular bounce in the model $Q=3H\delta\rho_{\rm de}$ in Section \ref{sec:bounce}. In both cases, the background evolution is realised in a flat FLRW universe with identical present-day cosmological parameters, namely $H_0=67.4~\mathrm{km\,s^{-1}\,Mpc^{-1}}$, $\Omega_{\rm (r,0)}=9\times10^{-5}$, $\Omega_{\rm (bm,0)}=0.05$, $\Omega_{\rm (dm,0)}=0.26$, $\Omega_{\rm (de,0)}=0.69$, and $w=-1$, differing only in the form of the dark-sector interaction.

Specifically, the crunch is realised in a two-fluid model when the interaction strength satisfies condition \eqref{eq:Q_dm_crunch_con}, as shown in Figure \ref{fig:crunch_limit} and derived in Appendix \ref{sec:mimmax}. The underlying mechanism is illustrated by the background evolution presented in Figure \ref{fig:crunch_5}. Strong energy transfer drives the dark energy density negative in the future, rendering it gravitationally attractive according to \eqref{eq:de_neg_attractive}. As a result, the universe decelerates and the violation of the strong energy condition required for accelerated expansion is removed. Once the total energy density satisfies $\rho_{\rm tot}=0$, the Hubble parameter vanishes, $H=0$, corresponding to a turnaround from expansion to contraction at a maximum scale factor given by \eqref{eq:Q_dm_a_max}. During the subsequent contraction phase, the direction of energy transfer reverses and dark energy becomes positive again. The scale factor then decreases rapidly until a future big crunch singularity is reached, characterised by $a\rightarrow 0$ and $\rho_{\rm tot}\rightarrow \infty$. This sequence of events is summarised step by step in Section \ref{sec:crunch_steps} and Table \ref{tab:crunch_phases_ide}.  Phase portraits are also provided in Figure \ref{fig:DSA_H_Qdm}, which show similar crunch-capable turnarounds for a larger selection of initial conditions, all of which correspond to the SiDEDM regime where dark energy become negative in the future, as seen in the left panel of \ref{fig:2D_Q_general_Phase_portrait_boundaries}.

A past non-singular bounce can likewise be realised in a two-fluid model when the coupling satisfies condition \eqref{eq:Q_dm_bounce_con}, as illustrated in Figure \ref{fig:bounce_limit} and also derived in Appendix \ref{sec:mimmax}. The corresponding mechanism is shown in Figure \ref{fig:bounce_5}. Evolving backward in time from the present, the universe contracts with $H<0$ and the direction of energy transfer reverses, causing energy to flow from dark matter to dark energy. This process drives the dark matter density negative, rendering it repulsive according to \eqref{eq:dm_neg_repulsive}. Once the negative dark matter component dominates, condition \eqref{BB.1} is satisfied, leading to $w^{\rm eff}_{\rm tot}<-1$ and the violation of both the strong and null energy conditions required for a non-singular bounce. The bounce occurs at a finite minimum scale factor $a_{\rm min}\neq0$ given by \eqref{eq:a_min_1}. Prior to the bounce, the universe contracts and the dark matter density returns to positive values. This evolution is summarised in Section \ref{sec:bounce_steps} and Table \ref{tab:bounce_phases_ide}. Additional phase portraits are provided in Figure \ref{fig:DSA_H_Qde}, which show similar bounce behaviour for a sets of initial conditions that have negative dark matter in the past, corresponding to the SiDEDM regime illustrated in the right panel of \ref{fig:2D_Q_general_Phase_portrait_boundaries}.

\begin{figure}
    \centering
    \includegraphics[width=0.9 \linewidth]{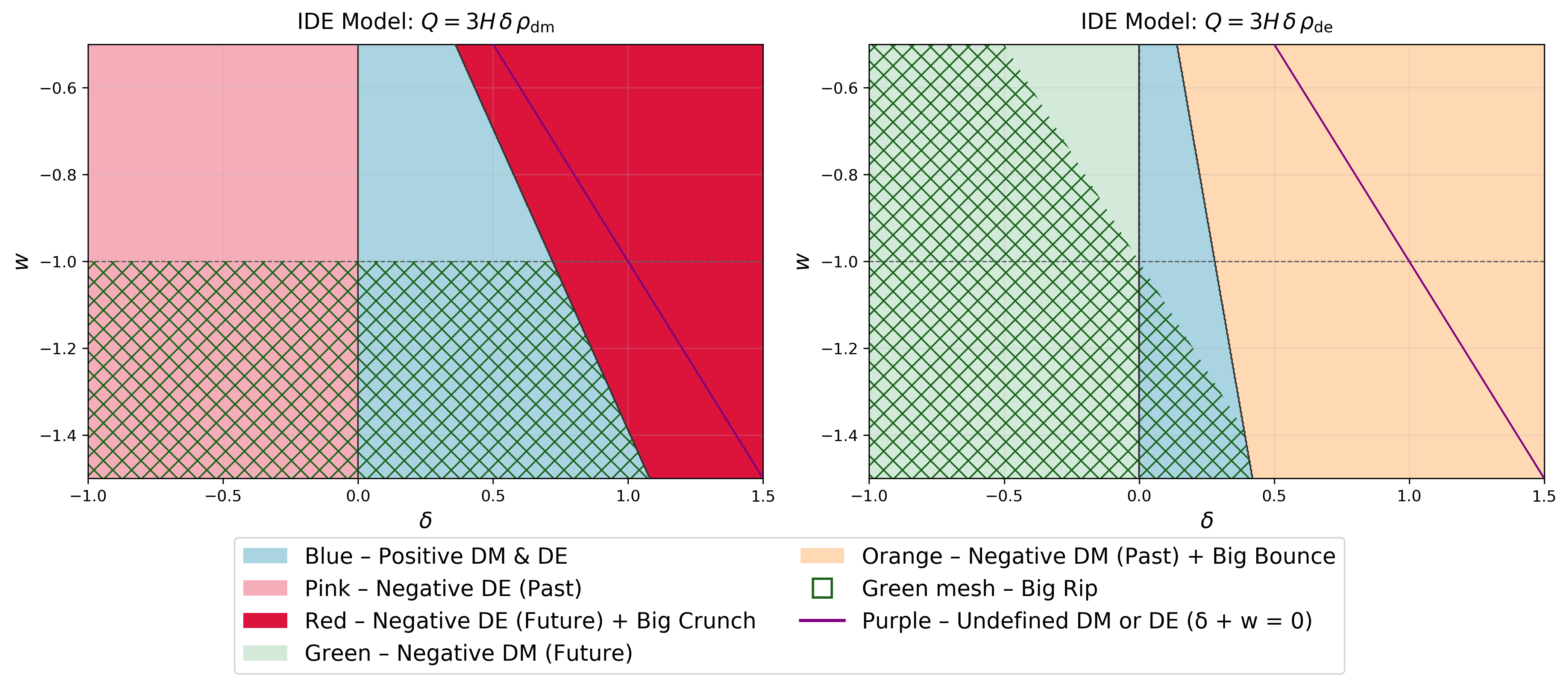}
    \caption{2D Portraits of the parameter space from Table \ref{tab:IDE_parameter_space_summary} for two IDE models. In both panels, blue areas indicate where the model has positive energy densities throughout all of cosmic evolution, purple shows undefined energy densities and the green mesh indicates future big rip singularities. For $Q= 3 H \delta \rho_{\text{dm}}$ (left panel), pink areas indicate negative DE in the past, while the red show negative DE in the future, which may lead to a future big crunch. For $Q= 3 H \delta \rho_{\text{de}}$ (right panel), green areas shows negative DM in the Future, while the orange area indicated negative DM in the past associated with a possible non-singular bounce. Scenarios where neither a big crunch, bounce or rip are mentioned, are predicted to have a past Big Bang singularity and eternal future expansion. Plots are created with $\Omega_{\rm{(dm,0)}}=0.266$ and $\Omega_{\rm{(de,0)}}=0.685$.}
    \label{fig:Q_parameter}
\end{figure}

These scenarios can be combined with the future big-rip solutions previously studied in \cite{vanderWesthuizen:2025I,vanderWesthuizen:2025III}, leading to the updated parameter-space classification for $Q=3H\delta\rho_{\rm dm}$ and $Q=3H\delta\rho_{\rm de}$ shown in Table \ref{tab:IDE_parameter_space_summary} and Figure \ref{fig:Q_parameter}. Related behaviour for other interaction kernels is discussed in Section \ref{sec:otherQ}, where both a crunch and a bounce are shown to be possible for $Q=3H\delta(\rho_{\rm dm}+\rho_{\rm de})$, depending on the initial conditions, as illustrated in Figures \ref{fig:crunch_bounce_limit} and \ref{fig:2D_Q_dm+de_Phase_portrait_boundaries}. Additionally, the inequalities derived in Appendix \ref{sec:mimmax} for five cases of $Q= 3 H (\delta_{\text{dm}} \rho_{\text{dm}} + \delta_{\text{de}}  \rho_{\text{de}})$ from the analytic expression for $a_{\rm min/max}$ are the existence conditions for a real, positive solution of $H=0$ at a turnaround.

\begin{table}[H]
\centering
\renewcommand{\arraystretch}{1.5}
\begin{tabular}{|c|c|c|}
\hline
\textbf{Scenario} 
& \multicolumn{2}{c|}{\textbf{Conditions for model}} \\
\cline{2-3}
& $\boldsymbol{Q = 3H\delta\rho_{\rm dm}}$ 
& $\boldsymbol{Q = 3H\delta\rho_{\rm de}}$ \\
\hline
\hline

Positive DM \& DE throughout
& $0 \leq \delta \leq -\frac{w}{(1 + r_0)}$ (weak DE$\rightarrow$DM transfer) 
& $0 \leq \delta \leq -\frac{w}{\left(1+\frac{1}{r_0}\right)}$ (weak DE$\rightarrow$DM transfer) \\
\hline

Negative DE (past)
& $\delta<0$ (DM$\rightarrow$DE in the past)
& -- \\
\hline

Negative DE (future) + Big Crunch
& $\delta>-\dfrac{w}{1+r_0}$ (strong DE$\rightarrow$DM transfer)
& -- \\
\hline

Negative DM (future)
& -- 
& $\delta<0$ (DM$\rightarrow$DE in the future) \\
\hline

Negative DM (past) + Big Bounce
& -- 
& $\delta>-\dfrac{w}{1+1/r_0}$ (strong DE$\rightarrow$DM transfer) \\
\hline

Future Big Rip Singularity
& Phantom regime ($w<-1$)
& Phantom regime ($\delta < -w-1 $) \\
\hline

Undefined DM or DE 
& $\delta + w = 0$ (divergent analytic solutions)
& $\delta + w = 0$ (divergent analytic solutions) \\
\hline

\end{tabular}
\caption{Summary of the cosmological scenarios shown in Figure \ref{fig:Q_parameter}, with the corresponding conditions for each interaction model. Here $r_0=\Omega_{(\rm dm,0)}/\Omega_{(\rm de,0)}$. Blank entries indicate scenarios that do not arise for the given interaction kernel. These scenarios are valid in the dark matter and dark energy two fluid description of the universe, while slight deviations are expected once a significant amount of baryons and radiation is present. Scenarios where neither a big crunch, bounce or rip are mentioned, are predicted to have a past Big Bang singularity and eternal future  expansion.}
\label{tab:IDE_parameter_space_summary}
\end{table}

A cyclic cosmological scenario can also be constructed for $Q=3H\delta\rho_{\rm dm}$ in a closed universe, as shown in Appendix \ref{sec:cyclic}. This case requires both spatial curvature and carefully chosen parameters and is therefore included primarily as a proof of principle rather than as a phenomenologically viable model.

One may ask whether the SiDEDM regime and other extreme regions of parameter space considered here are compatible with current cosmological observations. Background-level constraints using recent late-time probes were obtained in \cite{Figueruelo2026IDEconstraints}, which generally favour either energy transfer from dark matter to dark energy or a small transfer from dark energy to dark matter, rendering the scenarios discussed here disfavoured. Nevertheless, the SiDEDM regime for $Q=3H\delta\rho_{\rm de}$ remains allowed at the background level within both the 65\% and 95\% confidence intervals, and is therefore not excluded by current late-time data. Stronger conclusions will require the inclusion of early-time datasets and perturbation-level constraints, which we leave for future work.

Finally, we emphasise that the goal of this paper is not to replace standard cosmology or to propose immediately viable alternatives. Rather, our aim is twofold. First, we seek to better understand the structure of the parameter space of widely studied interacting dark energy models, demonstrating that exotic behaviours such as big crunches, non-singular bounces, and cyclic evolution can arise naturally in previously unexplored regimes. Importantly, these phenomena emerge at the background level within general relativity, without modifying gravity or introducing additional fields, but solely through dark-sector interactions that permit negative energy densities. This study may thus motivate others to search for similar dynamics in alternative interacting models, such as those already explored in \cite{Bruni:2021msx, Burkmar:2025svw}, or to explore the possibility of non-singular bounces in non-interacting models that may also feature negative dark energy or cosmological constants, such as those discussed in Section \ref{sec:neg_lit}. This work thus attempts to lay a foundation for more realistic cosmological models to be constructed in the future, which may have the potential to avoid theoretical concerns such as geodesic incompleteness at the Big Bang singularity.

Second, these results provide a framework for theoretically constraining interacting dark energy models. If negative energy densities or violations of the null energy condition are deemed physically unacceptable, the corresponding regions of parameter space may be excluded on theoretical grounds. This would restrict phenomenological IDE models to a smaller, better-behaved subset, such as the iDEDM regime characterised by weak energy transfer from dark energy to dark matter and strictly positive energy densities. The viability of this reduced parameter space should then be assessed using current and forthcoming cosmological datasets.

In conclusion, interacting dark energy models remain interesting candidates for addressing persistent tensions in cosmology and exhibit rich dynamics that warrant further theoretical and observational investigation.

\newpage

\appendix 

\section{Cyclic universe model case study: $(Q = 3H\delta\rho_{\text{dm}})$} \label{sec:cyclic}

\subsection{Mathematical background} \label{sec:math_cyc}

An additional possibility within interacting dark energy cosmology is the construction of a cyclic universe, in which cosmic evolution alternates between phases of expansion and contraction into both the past and the future. This appendix presents one explicit realization of such a scenario, obtained by combining a non-singular bounce with a subsequent big crunch. The model discussed here requires significant fine-tuning of parameters and is therefore unlikely to describe our universe. It is included primarily as a proof of principle, illustrating the range of background dynamics that can arise within the mathematical framework of dark sector interactions. The construction presented here is not unique, and other cyclic solutions with different properties may exist elsewhere in the IDE parameter space.

To construct a cyclic model, we focus on the interaction kernel $Q = 3H\delta\rho_{\rm dm}$ in a spatially closed universe. The inclusion of positive spatial curvature simplifies the construction of a cyclic evolution by allowing turnarounds to occur when the total energy density vanishes. As discussed throughout this work, both a turnaround preceding a big crunch and a non-singular bounce occur when the total energy density satisfies $\rho_{\rm tot}=0$. In the present case, we include curvature as an effective fluid with the standard evolution $\rho_{\rm k}(a)=\Omega_{\rm (k,0)}a^{-2}$, such that the general turnaround condition in \eqref{BCH.1} becomes
\begin{gather} \label{CC.1}
\begin{split}
\rho_{\rm tot} = \rho_{\rm r} + \rho_{\rm bm} + \rho_{\rm dm} + \rho_{\rm de} + \rho_{\rm k}
= 0 \quad \text{at turnaround } (a_{\rm min/max}) .
\end{split}
\end{gather}
For a cyclic model, different conditions must be satisfied to ensure a turnaround at $a_{\rm max}$ and a bounce at $a_{\rm min}$.

To obtain a turnaround at the maximum scale factor $a_{\rm max}$, strong deceleration is required, corresponding to the strong energy condition being satisfied. As in the big-crunch scenario discussed in Section \ref{sec:crunch}, this can be achieved through negative dark energy densities, $\rho_{\rm de}<0$. In a closed universe with $\rho_{\rm k}<0$, and neglecting radiation and baryons at large scale factors, condition \eqref{CC.1} reduces to
\begin{gather} \label{CC.2}
\begin{split}
\rho_{\rm tot} = \rho_{\rm dm} + \rho_{\rm de} + \rho_{\rm k}
= 0 \quad \text{at } a = a_{\rm max} , \\
\rightarrow \rho_{\rm dm} = -\rho_{\rm de} - \rho_{\rm k} .
\end{split}
\end{gather}

A non-singular bounce requires accelerated expansion, corresponding to violation of the strong energy condition, at the minimum scale factor $a_{\rm min}$. In the present model, this can be achieved through the negative pressure of a positive dark energy component with $\rho_{\rm de}>0$. The curvature contribution $\rho_{\rm k}<0$ ensures that condition \eqref{CC.1} is satisfied at the bounce, such that
\begin{gather} \label{CC.3}
\begin{split}
\rho_{\rm tot} = \rho_{\rm r} + \rho_{\rm bm} + \rho_{\rm dm} + \rho_{\rm de} + \rho_{\rm k}
= 0 \quad \text{at } a = a_{\rm min} , \\
\rightarrow \rho_{\rm r} + \rho_{\rm bm} + \rho_{\rm dm} + \rho_{\rm de}
= - \rho_{\rm k} .
\end{split}
\end{gather}

The maximum and minimum scale factors, $a_{\rm max}$ and $a_{\rm min}$, can be obtained by solving \eqref{CC.2} and \eqref{CC.3} numerically. Closed-form analytical solutions are difficult to obtain due to the presence of multiple fluid components.

Finally, we note that violation of the null energy condition is not required for the bounce in the curved universe considered here. The NEC violation condition derived in \eqref{BCdH.3} applies only to flat FLRW models, whereas \eqref{BCH.1} shows that a bounce can occur in a closed universe without invoking negative energy densities.

\subsection{Step-by-step cyclical model description} 

\begin{figure}
    \centering
    \includegraphics[width=0.85\linewidth]{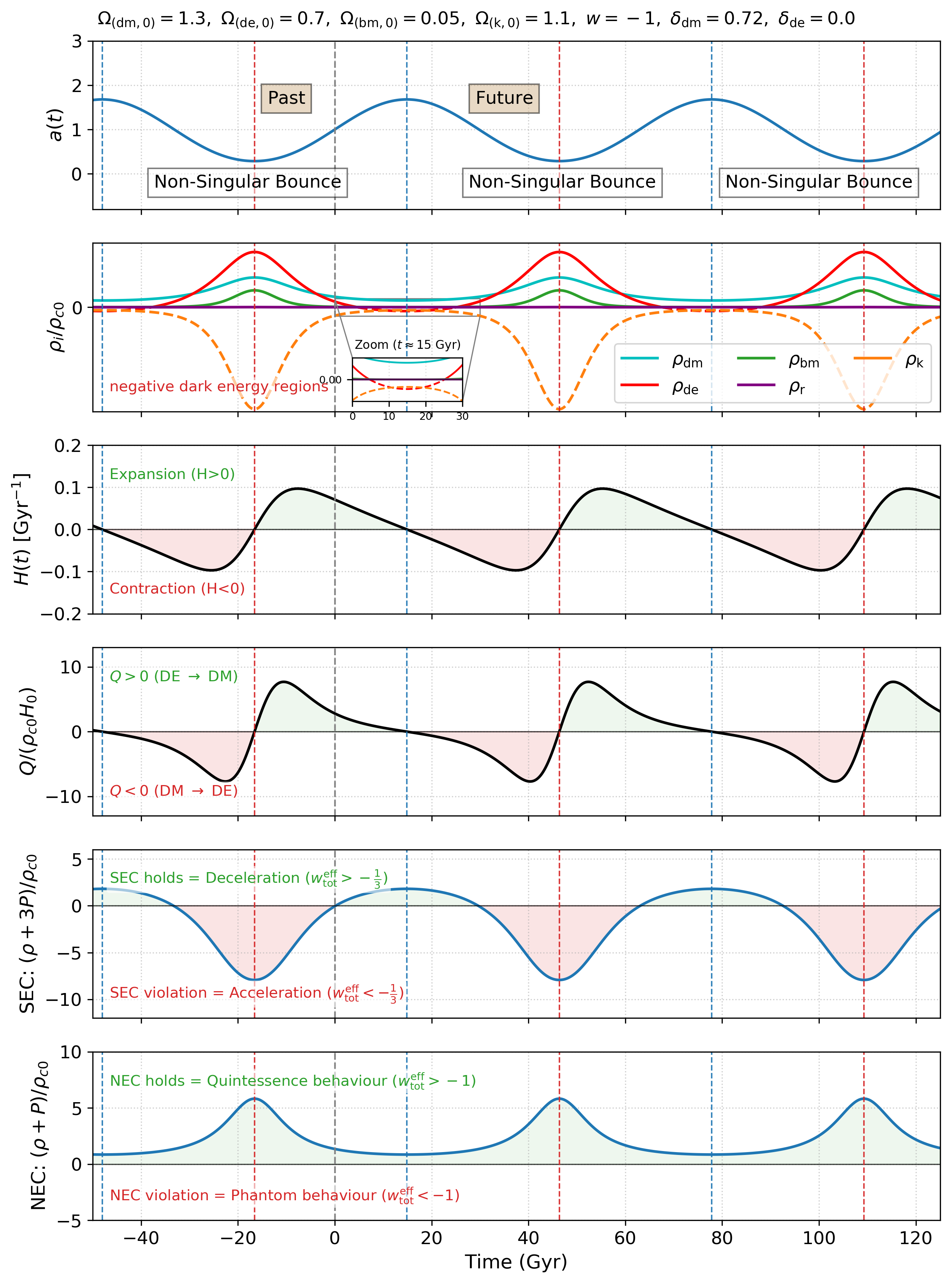}{}
    \caption{
    Evolution of the scale factor $a(t)$, dimensionless energy densities $\rho_i/\rho_{c0}$ (including the effective curvature contribution $\rho_{\rm k}$), Hubble parameter $H(t)$, interaction term $Q/(\rho_{c0}H_0)$, and the strong and null energy conditions (SEC and NEC) for a positively curved interacting dark energy model with interaction kernel $Q = 3H\delta\,\rho_{\rm dm}$. The present epoch ($a=1$) is characterised by accelerated expansion with $Q>0$. Moving forward in time, the interaction transfers energy from dark energy to dark matter, eventually rendering the dark energy density negative and attractive, as highlighted in the inset near the turnaround at $a_{\rm max}$. At this point the total energy density vanishes, $H=0$, and the expansion halts, leading to a turnaround into contraction. During the subsequent contracting phase the direction of energy transfer reverses ($Q<0$), while the NEC remains satisfied throughout due to the presence of positive spatial curvature. A non-singular bounce occurs at a minimum scale factor $a_{\rm min}$, where $H=0$ and the SEC is violated. This example illustrates a cyclic cosmological scenario enabled by interacting dark sectors.
}
    \label{fig:cyclic_5}
\end{figure}

We will now provide a step-by-step description of how a cyclical model can proceed, using the assumptions and set-up outlined in Section \ref{sec:math_cyc} of this Appendix. As previously mentioned, for this set-up we have chosen interaction kernel $Q= 3 H \delta \rho_{\text{dm}}$, with a strong energy transfer from dark energy to dark matter $(\delta_{\rm dm}=0.72)$,  alongside a universe with positive curvature $\Omega_{\rm (k,0)}=1.1$, which is realised by a significant increase in the present dark matter density $\Omega_{\rm (dm,0)}=1.3$. These parameter choices are mostly arbitrary, but were chosen for ease of illustrating the turnarounds described by \eqref{CC.2} and \eqref{CC.3}, and for the fact that they produce a period of accelerating expansion close to the present, both features seen in Figure \ref{fig:cyclic_5}. The structure will be similar to that outlined for the turnaround in Section \ref{sec:crunch_steps}, where we start from the present and move further into the future, mentioning interesting transitions that occur. These plots are once again made using the two Friedmann equations \eqref{F1} and \eqref{F2}, the evolution of the energy densities \eqref{eq:rho_dm_Q_dm} and \eqref{eq:rho_de_Q_dm}, and the energy conditions \eqref{1.3.1} and \eqref{BCdH.3}.

\begin{enumerate}
    \item \underline{Present accelerated expansion ($a=1 ;\;\rho_{\rm de}>0;\; H>0 ; \; Q>0 \;(\text{DE}\rightarrow\text{DM}); \; \text{SEC violated; NEC holds}$)}: \\ \\
    The model starts with initial conditions that describe a closed universe that underwent accelerated expansion with all energies positive, the SEC is violated and the NEC holds. The model includes a strong coupling $Q\gg0$ that causes energy transfer from dark enery to dark matter, leading to a period of decelerated expansion.
    \item \underline{Negative DE decelerated expansion ($1<a<a_{\rm{max}} ;\;\rho_{\rm de}<0;\; H>0 ; \; Q>0 \;(\text{DE}\rightarrow\text{DM}); \; \text{SEC \& NEC holds} $)}: \\ \\
    The large energy transfer from DE to DM causes the acceleration power of dark energy to diminish and the expansion to decelerate $w^{\rm{eff}}_{\rm{tot}}>-\frac{1}{3}$, so that the SEC holds. The energy transfer continues and dark energy becomes negative ($\rho_{\rm de}<0$) and gravitationally attractive as shown in \eqref{eq:de_neg_attractive}, which further decelerates the expansion rate.
    \item \underline{Turnaround at maximum scale factor ($a=a_{\rm{max}} ;\;\rho_{\rm de}<0;\; H=0 ; \; Q=0 \;(\text{no transfer}); \; \text{SEC \& NEC holds} $)}: \\ \\
    The dark energy component eventually becomes negative enough so that $\rho_{\rm tot}=0$, with a significant contribution from the curvature,  as given in condition \eqref{CC.2} and shown in the zoomed region around $t\approx15 Gyr$. From the Friedmann equation \eqref{F1}, this corresponds to $H=0$, where expansion stops at a maximum scaler factor $a_{\rm max}$. From the Raychaudhuri equation \eqref{F2}, we still have $\dot{H}\neq0$, as shown by the SEC that still holds, implying deceleration throughout the turnaround followed by a period of contraction. We note again that the total fluid effective equation of state description breaks down at the turnaround  ($w^{\rm eff}_{\rm tot} =\infty$).

    \item \underline{Negative DE decelerated contraction ($1<a<a_{\rm{max}} ;\;\rho_{\rm de}<0;\; H<0 ; \; Q<0 \;(\text{DM}\rightarrow\text{DE}); \; \text{SEC \& NEC holds} $)}: \\ \\
    As the universe begins to contract, the direction of energy transfer reverses (since $H<0$ and $Q \propto H$), such that energy flows from dark matter to dark energy. This energy transfer causes dark energy to become positive again, ending the period of decelerated contraction.
    
    \item \underline{Positive DE accelerated contraction ($a>a_{\rm{min}} ;\;\rho_{\rm de}>0;\; H<0 ; \; Q<0 \;(\text{DM}\rightarrow\text{DE}); \; \text{SEC violated; NEC holds}$)}: \\ \\
    In this phase, dark energy becomes positive enough to cause a slow down of the rate of contraction, corresponding to acceleration and SEC violation $w^{\rm{eff}}_{\rm{tot}}<-\frac{1}{3}$. In this specific case, acceleration will eventually cause the contraction to stop.

    \item \underline{Bounce at minimum scale factor ($a=a_{\rm{min}} ;\;\rho_{\rm de}>0;\; H=0 ; \; Q=0 \;(\text{no transfer}); \; \text{SEC violated; NEC holds}$)}: \\ \\
    The dominant dark energy component provides enouth acceleration another turnaround, or non-singular bounce at a minimum scale factor $a_{\rm min}\neq0$. At this bounce, the curvature component cancels out the positive energy densities of the other fluids such that condition \eqref{BCH.1} and \eqref{CC.3} holds. At the turnaround, we again have $H=0$, $Q=0$, $w^{\rm eff}_{\rm tot} =\infty$, while the SEC is violated ($\dot{H}>0$) to ensure the bounce is followed by a period of accelerated expansion, as described in step 1 above. Again, we note that all energies remain positive and that the NEC holds throughout the bounce, as the requirement for negative energies from \eqref{BCH.2} and the violation of the NEC from \eqref{BCdH.3}, is not required for a bounce in the presence of curvature.

\end{enumerate}

In Figure \ref{fig:bounce_5}, the six steps above repeat indefinitely into both the past and the future, allowing the model to avoid both past and future singularities.

\section{Finding $a_{\rm min/max}$ and turnaround existence conditions for different linear IDE kernels} \label{sec:mimmax}

In this appendix, provide expressions for the minimum and maximum scale factor that occurs at the turnarounds that occur at a bounce or before a crunch, respectively. The solution are obtained from the condition given in \eqref{BC.1}. The solutions provided here will only be valid given other conditions, such as those provided in Table \ref{tab:IDE_parameter_space_summary}, or other equivalent conditions for the other interactions considered below.

\subsection{Linear IDE model 1: \(Q= 3 H (\delta_{\text{dm}} \rho_{\text{dm}} + \delta_{\text{de}}  \rho_{\text{de}})\)} \label{Q_General}

Substituting expressions (2.2) and (2.3) for $\rho_{\rm{dm}}$ and $\rho_{\rm{de}}$ from \cite{vanderWesthuizen:2025III} into \eqref{BC.1}, we can solve for $a_{\rm{min/max}}$:
\begin{gather}
\begin{split} \label{eq:rhodm=de_ddm+dde}
 -& \frac{\delta_{\text{dm}}-\delta_{\text{de}}+w+\Delta}{4w \Delta} 
\Bigl[\rho_{\text{(de,0)}}\Bigl(\delta_{\text{dm}}-\delta_{\text{de}}+w-\Delta\Bigr)
+\rho_{\text{(dm,0)}}\Bigl(\delta_{\text{dm}}-\delta_{\text{de}}-w-\Delta\Bigr)\Bigr] 
a^{\frac{3}{2}\Bigl(\delta_{\text{dm}}-\delta_{\text{de}}-w-2+\Delta\Bigr)} \\[1mm]
+& \frac{\delta_{\text{dm}}-\delta_{\text{de}}+w-\Delta}{4w \Delta} 
\Bigl[\rho_{\text{(de,0)}}\Bigl(\delta_{\text{dm}}-\delta_{\text{de}}+w+\Delta\Bigr)
+\rho_{\text{(dm,0)}}\Bigl(\delta_{\text{dm}}-\delta_{\text{de}}-w+\Delta\Bigr)\Bigr] 
a^{\frac{3}{2}\Bigl(\delta_{\text{dm}}-\delta_{\text{de}}-w-2-\Delta\Bigr)}  \\
 = -& \frac{\delta_{\text{dm}}-\delta_{\text{de}}-w+\Delta}{4w \Delta} 
\Bigl[\rho_{\text{(de,0)}}\Bigl(\delta_{\text{dm}}-\delta_{\text{de}}+w-\Delta\Bigr)
+\rho_{\text{(dm,0)}}\Bigl(\delta_{\text{dm}}-\delta_{\text{de}}-w-\Delta\Bigr)\Bigr] 
a^{\frac{3}{2}\Bigl(\delta_{\text{dm}}-\delta_{\text{de}}-w-2+\Delta\Bigr)} \\[1mm]
+& \frac{\delta_{\text{dm}}-\delta_{\text{de}}-w-\Delta}{4w \Delta} 
\Bigl[\rho_{\text{(de,0)}}\Bigl(\delta_{\text{dm}}-\delta_{\text{de}}+w+\Delta\Bigr)
+\rho_{\text{(dm,0)}}\Bigl(\delta_{\text{dm}}-\delta_{\text{de}}-w+\Delta\Bigr)\Bigr] 
a^{\frac{3}{2}\Bigl(\delta_{\text{dm}}-\delta_{\text{de}}-w-2-\Delta\Bigr)}  \\
a_{\rm min/max} = \quad & \left[
\frac{
\left[ \Omega_{\text{(de,0)}} (\delta_{\text{dm}} - \delta_{\text{de}} + w + \Delta)
+ \Omega_{\text{(dm,0)}} (\delta_{\text{dm}} - \delta_{\text{de}} - w + \Delta) \right]
}{
\left[ \Omega_{\text{(de,0)}} (\delta_{\text{dm}} - \delta_{\text{de}} + w - \Delta)
+ \Omega_{\text{(dm,0)}} (\delta_{\text{dm}} - \delta_{\text{de}} - w - \Delta) \right]
}
\right]^{\frac{1}{3\Delta}} .
\end{split}   
\end{gather}

where $\Delta$ is the determinant:
\begin{gather}
\begin{split} \label{eq:determinant_Q_linear}
\Delta = \sqrt{(\delta_{\text{dm}}+\delta_{\text{de}}+w)^2 - 4\delta_{\text{de}}\delta_{\text{dm}}}= \sqrt{ \delta_{\text{dm}}^2 + \delta_{\text{de}}^2 + w^2 - 2 \delta_{\text{dm}} \delta_{\text{de}} + 2 \delta_{\text{dm}} w + 2 \delta_{\text{de}} w }\,. 
\end{split}   
\end{gather}
In order for $a_{\rm min\max}$ to be both real and positive, we need the denominator and numerator in the bracket base of \eqref{eq:rhodm=de_ddm+dde} to be the same sign. It can be shown that after substituting \eqref{eq:determinant_Q_linear} into \eqref{eq:rhodm=de_ddm+dde}, assuming $w<0$ and simplifying algebraically, we obtain the conditions in \eqref{DSA.Q.delta_dm+delta_de.PEC_BG} that allows this.
\begin{gather} \label{DSA.Q.delta_dm+delta_de.PEC_BG}
\text{Bounce/crunch existence: } 1.  \;  \delta_{\text{dm}} r_0 + \delta_{\text{de}}> -\dfrac{w r_0}{1+r_0};\quad 
2.\; (\delta_{\text{dm}}+\delta_{\text{de}}+w)^2 \ge 4\,\delta_{\text{de}}\delta_{\text{dm}}.
\end{gather}
In \eqref{DSA.Q.delta_dm+delta_de.PEC_BG}, condition $1$ corresponds to the SiDEDM regime and a violation of the previously derived positive energy condition in \cite{vanderWesthuizen:2025I} , which ensures negative energies required for $\rho_{\rm tot}=0$ at some point (the same holds for the equivalent conditions derived in the four special cases of this model). Condition $2$  is to ensure that the determinant $\Delta$ in \eqref{eq:determinant_Q_linear} remains real.

\subsection{Linear IDE model 2: \(Q= 3 H \delta(\rho_{\text{dm}} + \rho_{\text{de}})\)} \label{Q_dm+de}

Setting $\delta_{\rm dm}=\delta_{\rm de}=\delta$ in \eqref{eq:rhodm=de_ddm+dde} and \eqref{eq:determinant_Q_linear}, gives:
 
\begin{equation}
\begin{split}
a_{\rm min/max}
= \left[ 
\frac{
\Omega_{\text{(de,0)}} ( w + \Delta) + \Omega_{\text{(dm,0)}} ( - w + \Delta)
}{
\Omega_{\text{(de,0)}} ( w - \Delta) + \Omega_{\text{(dm,0)}} ( - w - \Delta)
}
\right]^{\frac{1}{3 \Delta}},
\end{split}
\label{eq:a_min_max_dm+de_A}
\end{equation}
where $\Delta = \sqrt{w(4\delta + w)}$. Similarly, we can derive the existence condition a positive and real $a_{\rm min/max}$ from \eqref{eq:a_min_max_dm+de_A}, or by setting  $\delta_{\rm dm}=\delta_{\rm de}=\delta$ in \eqref{DSA.Q.delta_dm+delta_de.PEC_BG}, both cases leading to \eqref{DSA.Q.dm+de.PEC_BG}.

\begin{gather} \label{DSA.Q.dm+de.PEC_BG}
\text{Bounce/crunch existence: } \;1. \;\delta > -\frac{w r_0}{(1+r_0)^2}  \quad ;  \quad 2. \; \delta \le -\frac{w}{4}.
\end{gather}
Condition $1$ ensures negative energy and condition $2$ guarantees real values.

\subsection{Linear IDE model 3: \(Q= 3 H \delta(\rho_{\text{dm}} - \rho_{\text{de}})\)} \label{Q_dm-de}

Setting $\delta_{\rm dm}=\delta$ and $\delta_{\rm de}=-\delta$ in \eqref{eq:rhodm=de_ddm+dde} and \eqref{eq:determinant_Q_linear}, gives:

\begin{equation}
\begin{split}
a_{\rm min/max} &= 
\left[\frac{\left[\Omega_{\text{(de,0)}} (2\delta + w + \Delta) + \Omega_{\text{(dm,0)}} (2\delta - w + \Delta) \right]}{ \left[\Omega_{\text{(de,0)}} (2\delta + w - \Delta) + \Omega_{\text{(dm,0)}} (2\delta - w - \Delta) \right]}\right]^{\frac{1}{3\Delta}},
\end{split}
\label{eq:rhodm=de_dm-de}
\end{equation}
where $\Delta = \sqrt{4\delta^2 + w^2}$ and is always real. A positive and real $a_{\rm min/max}$ can be obtained from \eqref{eq:rhodm=de_dm-de}, or by setting $\delta_{\rm dm}=\delta$ and $\delta_{\rm de}=-\delta$ in \eqref{DSA.Q.delta_dm+delta_de.PEC_BG}, both cases leading to \eqref{DSA.Q.dm-de.PEC_BG}.
\begin{gather} \label{DSA.Q.dm-de.PEC_BG}
\text{Bounce/crunch existence: } \;1. \;\text{ if } \; r_0<1 \text{ then } \;\delta < \frac{w r_0}{1-r_0^2} \quad ; \;2. \;\text{ if } \; r_0>1 \text{ then } \;\delta > \frac{w r_0}{1-r_0^2}.
\end{gather}

\subsection{Linear IDE model 4: \(Q= 3 H \delta \rho_{\text{dm}}\)} \label{Q_dm}

This model will always have positive dark matter, as seen in \eqref{eq:rho_dm_Q_dm}.  From condition \eqref{BCdH.5}, any turnaround will therefore be associated with a crunch at $a_{\rm max}$. Setting $\delta_{\rm dm}=\delta$ and $\delta_{\rm de}=0$ in \eqref{eq:rhodm=de_ddm+dde} and \eqref{eq:determinant_Q_linear}, and simplifying gives:

\begin{equation}
\begin{split}
a_{\rm max}  =& \left[ -\frac{\Omega_{\text{(de,0)}}}{\Omega_{\text{(dm,0)}}} \left(\frac{\delta+w}{w} \right) -\frac{\delta}{w}   \right]^{\frac{1}{{3(\delta +w) }}}. \\
\end{split}
\label{eq:rhodm=de_dm}
\end{equation}
 A positive and real $a_{\rm max}$ can be obtained by requiring a positive base in \eqref{eq:rhodm=de_dm} with $w<-1$, or by setting $\delta_{\rm dm}=\delta$ and $\delta_{\rm de}=0$ in \eqref{DSA.Q.delta_dm+delta_de.PEC_BG}, both cases leading to \eqref{eq:rhodm=de_dm_2}.
\begin{equation}
\begin{split}
\text{Crunch existence: } \;\delta >-\frac{w}{1+r_0}. \\
\end{split}
\label{eq:rhodm=de_dm_2}
\end{equation}
Condition \eqref{eq:rhodm=de_dm_2} violates the upper positive energy condition derived in \cite{vanderWesthuizen:2023hcl, vanderWesthuizen:2025I}.
\subsection{Linear IDE model 5: \(Q= 3 H \delta\rho_{\text{de}}\)} \label{Q_de}
This model will always have positive dark energy, as seen in \eqref{eq:rho_de_Q_de}. From condition \eqref{BCdH.5}, any turnaround will therefore be associated with a bounce at $a_{\rm min}$.
Setting $\delta_{\rm dm}=0$ and $\delta_{\rm de}=\delta$ in \eqref{eq:rhodm=de_ddm+dde} and \eqref{eq:determinant_Q_linear}, and simplifying gives:

\begin{equation}
\begin{split}
a_{\rm min}  =& \left[-\frac{   \Omega_{\text{(dm,0)}} (  \delta+w)  }{ w\Omega_{\text{(de,0)}}  } -\frac{\delta}{w}\right]^{-\frac{1}{{3(\delta +w) }}}. \\
\end{split}
\label{eq:rhodm=de_dm_Qde}
\end{equation}
Alternatively, to include baryons, we substitute $\rho_{\rm bm}=\rho_{\text{(bm,0)}}a^{-3}$ and the energy densities \eqref{eq:rho_dm_Q_de} and \eqref{eq:rho_de_Q_de} into the following requirement:
\begin{gather}
\begin{split} \label{eq:a_min_bm_Q_de}
\rho_{\rm{bm}}+\rho_{\rm{de}}&= -\rho_{\rm{dm}} \\
\rho_{\text{(bm,0)}}a^{-3}+\rho_{\text{(de,0)}} a^{-3(\delta + w + 1)}&=-  \left(\rho_{\text{(dm,0)}}  + \rho_{\text{(de,0)}} \left(\frac{\delta}{\delta+w} \right) \left[1-  a^{-3(\delta + w)} \right] \right) a^{-3} \\
        a_{\rm min}&= \left[- \frac{\rho_{\text{(bm,0)}}+  \rho_{\text{(dm,0)}}}{\rho_{\text{(de,0)}}} \left(\frac{\delta+w}{w} \right) - \left(\frac{\delta}{w} \right) \right]^{\frac{1}{-3(\delta + w )}}.   \\
\end{split}   
\end{gather}

 A positive and real $a_{\rm min}$ can be obtained by requiring a positive base in \eqref{eq:rhodm=de_dm_Qde} with $w<-1$, or by setting $\delta_{\rm dm}=0$ and $\delta_{\rm de}=\delta$ in \eqref{DSA.Q.delta_dm+delta_de.PEC_BG}, both cases leading to \eqref{eq:rhodm=de_dm_Qde2}.
\begin{equation}
\begin{split}
\text{Bounce existence: } \;\delta >-\frac{w}{1+\frac{1}{r_0}}. \\
\end{split}
\label{eq:rhodm=de_dm_Qde2}
\end{equation}
Condition \eqref{eq:rhodm=de_dm_Qde2} violates the upper positive energy condition derived in \cite{vanderWesthuizen:2023hcl, vanderWesthuizen:2025I}.

\section{Phase portraits showcasing negative energy crossing in SiDEDM regime} \label{sec:DSA}

\begin{figure}[htbp]
    \centering
    \begin{subfigure}[b]{0.49\linewidth}
        \centering
        \includegraphics[width=\linewidth]{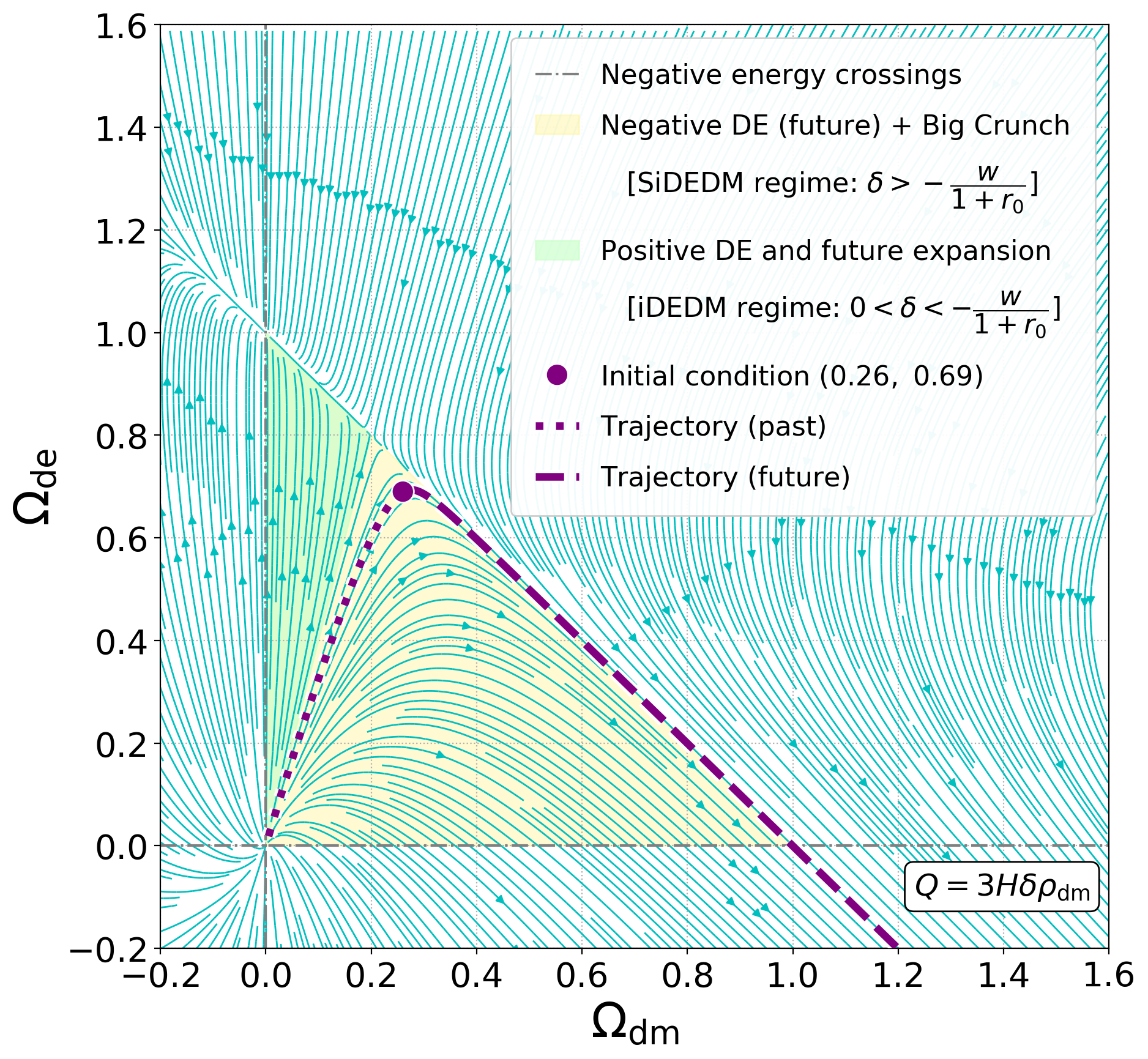}
        \label{fig:iDEDM2D}
    \end{subfigure}
    \hspace{0pt} 
    \begin{subfigure}[b]{0.49\linewidth}
        \centering
        \includegraphics[width=\linewidth]{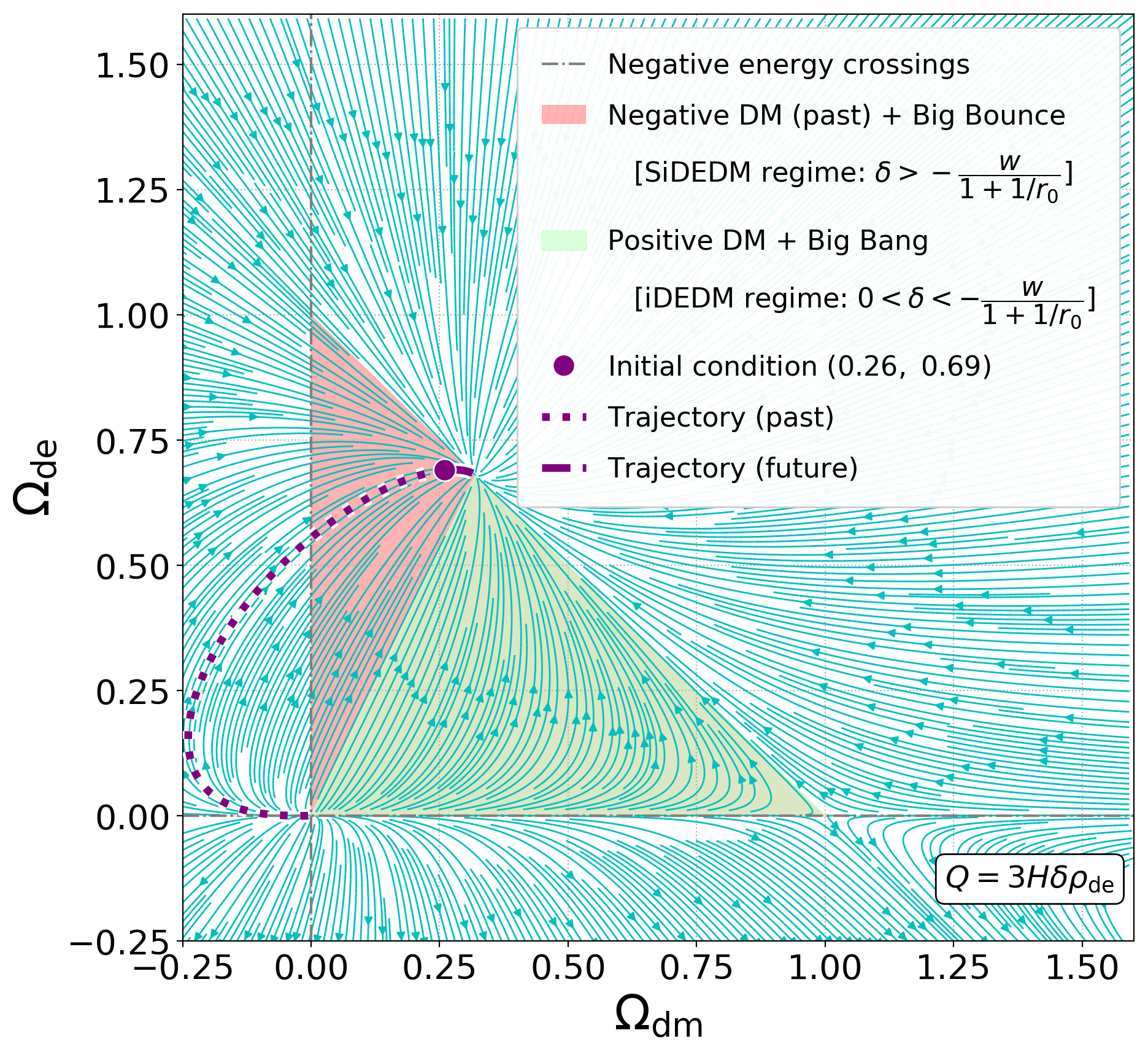}
        \label{fig:iDMDE2D}
    \end{subfigure}
\caption{2D projection of the phase portrait for $Q= 3 H \delta \rho_{\text{dm}}$ (left panel, with $\delta=0.80$) and $Q= 3 H \delta \rho_{\text{de}}$ (right panel, with $\delta=0.32$), showing positive-energy trajectories in the iDEDM regime, but negative energies in the SiDEDM regime. Interaction $Q= 3 H \delta \rho_{\text{dm}}$ can never have negative dark matter densities, as energy transfer stops $Q=0$ when $\rho_{\rm dm}=0$, which corresponds here to an invariant sub-manifold at $\Omega_{\rm dm}=0$. Negative dark energies are allowed in the future which may lead to a big crunch. Similarly, interaction $Q= 3 H \delta \rho_{\text{de}}$ has an invariant sub-manifold at $\Omega_{\rm de}=0$, since $Q=0$ at $\rho_{\rm{ de}}=0$, and will therefore always have positive dark energy densities. Instead, dark matter may become negative which allows non-singular bounce. 
A visualization of what happens in a given volume element of space to create these negative energies is provided in Figure \ref{fig:neg_dm_visual} for $Q= 3 H \delta \rho_{\text{dm}}$, and Figure \ref{fig:neg_de_visual} for $Q= 3 H \delta \rho_{\text{de}}$. Visualizations of how negative dark energy in the future or negative dark energy in the past lead to either a big crunch or non-singular bounce is illustrated in Figure \ref{fig:crunch_limit} and \ref{fig:bounce_limit}, respectively.}
    \label{fig:2D_Q_general_Phase_portrait_boundaries}
\end{figure}

\begin{figure}
    \centering
    \includegraphics[width=0.9 \linewidth]{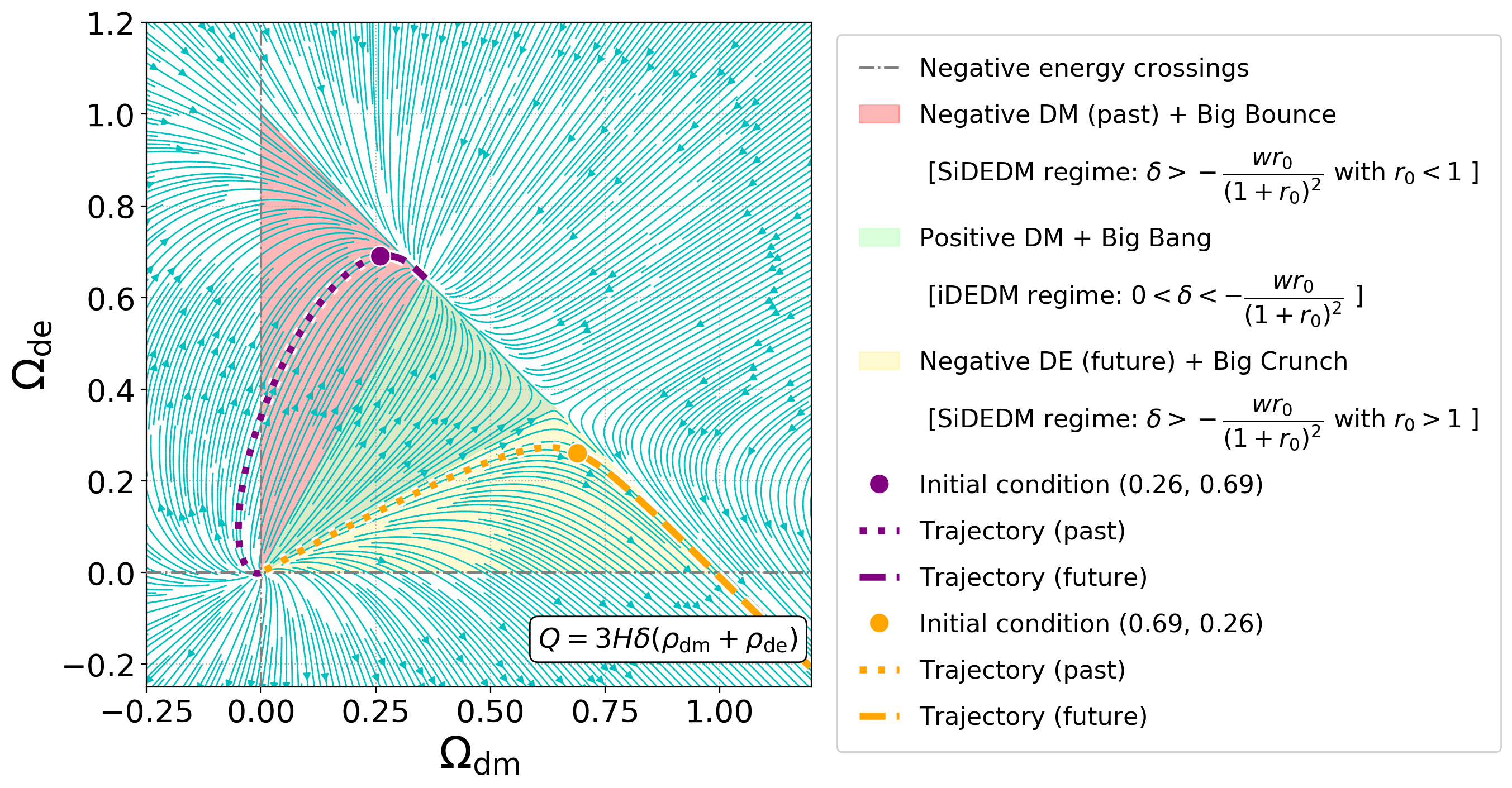}
    \caption{2D projection of the phase portrait for $Q= 3 H \delta (\rho_{\text{dm}}+\rho_{\text{de}})$, with $\delta=0.23$, showing positive-energy trajectories in the iDEDM regime, but negative energies in the SiDEDM regime. This interaction can have both negative dark matter and dark energy densities, as energy transfer does not stop $Q\neq0$ when $\rho_{\rm dm}=0$ or $\rho_{\rm de}=0$, which corresponds here to the absence of invariant sub-manifold at $\Omega_{\rm dm}=0$ and $\Omega_{\rm de}=0$. Depending on the initial conditions in the SiDEDM regime, trajectories will have either past negative dark matter densities alongside a bounce, or alternatively negative dark energy in the future, and hence a big crunch. These scenarios are illustrated in in Figure \ref{fig:crunch_bounce_limit}, showing a big crunch (if the ratio $r_0>1$) or non-singular bounce (if $r_0<1$), in the SiDEDM regime.}
    \label{fig:2D_Q_dm+de_Phase_portrait_boundaries}
\end{figure}

We consider the same system of equations studied in \cite{vanderWesthuizen:2025I}, which describes the evolution of dark energy ${\Omega}'_{\rm{de}}$, dark matter ${\Omega}'_{\rm{dm}}$ and baryonic matter ${\Omega}'_{\rm{de}}$ given in \eqref{DSA.9}. 

\begin{gather} \label{DSA.9}
\begin{split}
{\Omega}'_{\rm{de}} &= \Omega_{\rm{de}} \left[ 1 - \Omega_{\rm{bm}} - \Omega_{\rm{dm}} - \Omega_{\rm{de}}  \left(1 - 3 w \right) - 3w \right]  - \frac{8 \pi G }{3H^3} Q, \\
{\Omega}'_{\rm{dm}} &= \Omega_{\rm{dm}} \left[ 1 - \Omega_{\rm{bm}} - \Omega_{\rm{dm}} - \Omega_{\rm{de}}  \left(1 - 3 w \right) \right]  + \frac{8 \pi G }{3H^3} Q, \\
{\Omega}'_{\rm{bm}} &= \Omega_{\rm{bm}} \left[ 1 - \Omega_{\rm{bm}} - \Omega_{\rm{dm}} - \Omega_{\rm{de}}  \left(1 - 3 w \right) \right].
\end{split}
\end{gather}
The derivative here is with regard to the Hubble parameter ${\Omega}'_{\rm{i}} = \frac{d}{H \, dt} \Omega_{\rm{i}}$. Since $\Omega$ is undefined at $H=0$, the trajectories will therefore not be defined at the turnaround, but can show how negative energies develop prior to this point. The flatness assumption is included in \eqref{DSA.9}, such that radiation density is obtained from  $\Omega_{\rm{r}} = 1 - \Omega_{\rm{bm}} - \Omega_{\rm{dm}} - \Omega_{\rm{de}}$. Using \eqref{DSA.9}, we may obtain the 2D phase portraits in the $(\Omega_{\rm{dm}},\Omega_{\rm{de}})$ plane by setting ${\Omega}_{\rm{bm}}=0$. Substituting $Q= 3 H \delta \rho_{\text{dm}}$ and  $Q= 3 H \delta \rho_{\text{de}}$ into \eqref{DSA.9}, we obtain the 2D phase portraits in Figure \ref{fig:2D_Q_general_Phase_portrait_boundaries}, while the phase portrait for $Q= 3 H \delta (\rho_{\text{dm}}+\rho_{\text{de}})$ is given in Figure \ref{fig:2D_Q_dm+de_Phase_portrait_boundaries}. A full dynamical system analysis for the four fluid system, including critical points and stability conditions, is given in \cite{vanderWesthuizen:2025I, vanDerWesthuizen:2025SAIP}. There it was shown that invariant sub-manifolds exist (also present in Figure \ref{fig:2D_Q_general_Phase_portrait_boundaries}) at $\Omega_{\rm dm}=0$ for $Q= 3 H \delta \rho_{\text{dm}}$, and at $\Omega_{\rm de}=0$ for $Q= 3 H \delta \rho_{\text{de}}$, which corresponds to a stop of energy transfer $Q=0$ when the corresponding dark fluid is depleted $\rho_{\rm{dm/de}}=0$, hence preventing negative energies arising for that fluid. Conversely, negative dark matter may arise for $Q= 3 H \delta \rho_{\text{de}}$ and negative dark energy for $Q= 3 H \delta \rho_{\text{dm}}$, while both can arise for $Q= 3 H \delta (\rho_{\text{dm}}+\rho_{\text{de}})$. These scenarios may lead to crunching or bouncing scenarios, as discussed in Section \ref{sec:crunch} and \ref{sec:bounce}. The parameters chosen for $\delta$ in Figure \ref{fig:2D_Q_general_Phase_portrait_boundaries} and \ref{fig:2D_Q_dm+de_Phase_portrait_boundaries} have been chosen to approximately match the cases in Figure \ref{fig:crunch_limit}, \ref{fig:bounce_limit} and \ref{fig:crunch_bounce_limit}, while small changes were made to account for the 2 fluid nature of these portraits and to ease illustration. For other dynamical systems analysis of similar linear interacting dark energy models, but using a different system of equations from \eqref{DSA.9}, as well as discussions on negative energies in these models, see \cite{Valiviita:2008iv, Quartin_2008, He:2008tn, He:2008si, Caldera_Cabral_2009_DSA, Caldera_Cabral_2009_structure, Izquierdo_2018, Pan_2020, Panotopoulos_2020, Deogharia_2021, von_Marttens_2020, Ar_valo_2022}.

\section{Derivation of dynamical system showcasing turnarounds associated with negative energies} \label{sec:DSA_new}
In order to construct a dynamical system that shows flow lines through the point $H=0$, we need to define new parameters that exist at this point and show the evolution of the Hubble parameter $H$. This can be achieved by dividing the densities by the critical density $\rho_{(c,0)}=\frac{3H_0^2}{8 \pi G}$. We choose the following dimensionless variables in a flat two fluid system:
\begin{gather} \label{DSA2.1}
\begin{split}
x_{\rm dm}=\frac{\rho_{\rm{dm}}}{\rho_{(c,0)}} ; \quad x_{\rm de}=\frac{\rho_{\rm{de}}}{\rho_{(c,0)}} ; \quad h=\frac{H}{H_0}.
\end{split}
\end{gather}
Since we assume flatness, we also have the constraint $h^2=x_{\rm dm}^2+x_{\rm de}^2$. We will use the derivative with regard to dimensionless time variable $\tau=H_0t$, such that we have:
\begin{gather} \label{DSA2.2}
\begin{split}
 \frac{d}{d \tau} = \frac{1}{H_0}\frac{d}{dt} \quad \rightarrow \quad   \frac{d}{dt}=H_0\frac{d}{d \tau}. 
\end{split}
\end{gather}
We obtain our system of equations by dividing the conservation equations \eqref{eq:conservation.1} by $\rho_{(c,0)}$ and transforming the derivatives to be with regard to $\tau$
\begin{gather} \label{DSA2.3}
\begin{split}
\frac{H_0 }{\rho_{(c,0)}}\frac{d \rho_{\text{dm}}}{d \tau}  +  \frac{3H }{\rho_{(c,0)}}\rho_{\text{dm}} = \frac{Q }{\rho_{(c,0)}}\; \quad &\rightarrow \quad  \frac{d x_{\text{dm}}}{d \tau}   = -3hx_{\text{dm}} +\frac{Q }{H_0\rho_{(c,0)}}
\;,\\
\frac{H_0 }{\rho_{(c,0)}}\frac{d \rho_{\text{de}}}{d \tau}  +  \frac{3H }{\rho_{(c,0)}}\rho_{\text{de}}(1+w) = -\frac{Q }{\rho_{(c,0)}}\; \quad& \rightarrow \quad  \frac{d x_{\text{de}}}{d \tau}   = -3h(1+w)x_{\text{de}} -\frac{Q }{H_0\rho_{(c,0)}}.
 \end{split}
\end{gather}
Substituting in the interaction kernel $Q= 3 H (\delta_{\text{dm}} \rho_{\text{dm}} + \delta_{\text{de}}  \rho_{\text{de}})$ into \eqref{DSA2.3}, gives:
\begin{gather} \label{DSA2.4}
\begin{split}
  \frac{d x_{\text{dm}}}{d \tau}   = -3hx_{\text{dm}} +\frac{ 3 H (\delta_{\text{dm}} \rho_{\text{dm}} + \delta_{\text{de}}  \rho_{\text{de}}) }{H_0\rho_{(c,0)}} \quad& \rightarrow \quad  \frac{d x_{\text{dm}}}{d \tau}   = 3h\left[ (\delta_{\text{dm}}-1) x_{\text{dm}} + \delta_{\text{de}}  x_{\text{de}}\right],\\
   \frac{d x_{\text{de}}}{d \tau}   = -3hx_{\text{de}}(1+w) -\frac{ 3 H (\delta_{\text{dm}} \rho_{\text{dm}} + \delta_{\text{de}}  \rho_{\text{de}}) }{H_0\rho_{(c,0)}} \quad& \rightarrow \quad  \frac{d x_{\text{de}}}{d \tau}   = -3h\left[ \delta_{\text{dm}} x_{\text{dm}} + (1+w+\delta_{\text{de}} ) x_{\text{de}}\right].
 \end{split}
\end{gather}
The final equation is derived from the Friedmann equation \eqref{F1} and Raychaudhuri equation \eqref{F2} in a flat universe, which can be rewritten in the following form:
\begin{gather} \label{DSA2.5}
\begin{split}
\dot{H}  & = - 4\pi G\left[\rho_{\rm dm} + \rho_{\rm de} + P_{\rm de} \right]=- 4\pi G\left[\rho_{\rm dm} + (1+w)\rho_{\rm de} \right]=-\frac{3}{2}H_0^2\left[x_{\rm dm} + (1+w)x_{\rm de} \right],  
\end{split}
\end{gather} 
where in the last step we used the relation $4\pi G\rho_{\rm (c,0)}=\frac{3}{2}H_0^2$.
Converting the derivative in \eqref{DSA2.5} to $\tau$ using \eqref{DSA2.2}, gives:
\begin{gather} \label{DSA2.6}
\begin{split}
 \frac{dh}{d \tau}  & =-\frac{3}{2}\left[x_{\rm dm} + (1+w)x_{\rm de} \right].  \\
\end{split}
\end{gather} 
Combining \eqref{DSA2.4} and \eqref{DSA2.6}, leads to the system of equations:
\begin{gather} \label{DSA2.7}
\begin{split}
 \frac{d x_{\text{dm}}}{d \tau}   = 3h\left[ (\delta_{\text{dm}}-1) x_{\text{dm}} + \delta_{\text{de}}  x_{\text{de}}\right]; \quad    \frac{d x_{\text{de}}}{d \tau}   = -3h\left[ \delta_{\text{dm}} x_{\text{dm}} + (1+w+\delta_{\text{de}} ) x_{\text{de}}\right]; \quad    \frac{dh}{d \tau}  & =-\frac{3}{2}\left[x_{\rm dm} + (1+w)x_{\rm de} \right] .  \\
\end{split}
\end{gather}
For the simplified case where $w=-1$ used in all phase portraits, this system of equations reduces to:
\begin{gather} \label{DSA2.8}
\begin{split}
 \frac{d x_{\text{dm}}}{d \tau}   = 3h\left[ (\delta_{\text{dm}}-1) x_{\text{dm}} + \delta_{\text{de}}  x_{\text{de}}\right]; \quad   \frac{d x_{\text{de}}}{d \tau}   = -3h\left[ \delta_{\text{dm}} x_{\text{dm}} + \delta_{\text{de}}  x_{\text{de}}\right]; \quad   \frac{dh}{d \tau}   =-\frac{3}{2}x_{\rm dm} .  \\
\end{split}
\end{gather} 
For a turnaround at $h=0$ to be a bounce, we need $\frac{dh}{d \tau}>0$, and from \eqref{DSA2.7}, we can see this corresponds to $x_{\rm dm}<0$ or equivalently $\rho_{\rm dm}<0$. Similarly, for a turnaround preceding a big crunch, we require $\frac{dh}{d \tau}<0$ and thus $
x_{\rm dm}>0$ and $\rho_{\rm dm}>0$ at $h=0$. These arguments correspond to the condition derived in \eqref{BCdH.5}.

\bibliographystyle{apsrev4-2}
\bibliography{References, biblio, References_bounce}  

\end{document}